\pdfoutput=1
\documentclass[aps,11pt,a4paper,pre,byrevtex,bibnotes,nofootinbib,showpacs,notitlepage,showkeys,longbibliography]{revtex4-1}
\usepackage{microtype}
\usepackage{graphicx}
\usepackage{amsmath}
\usepackage{amssymb}
\usepackage{amsthm}
\usepackage{color}
\definecolor{myblue}{rgb}{0.153,0.322,0.706}
\usepackage[colorlinks,linkcolor=myblue,urlcolor=myblue,citecolor=myblue,breaklinks=true]{hyperref}

\newcommand{\be}{\begin{equation}}
\newcommand{\ee}{\end{equation}}
\newcommand{\ra}{\rightarrow}
\newcommand{\p}{\partial}
\newcommand{\transp}{\mathsf{T}}
\newcommand{\Dt}{\Delta t}
\newcommand{\Z}{\mathbb{Z}}
\newcommand{\N}{\mathbb{N}}
\newcommand{\R}{\mathbb{R}}
\newcommand{\E}{\mathbb{E}}
\newcommand{\proba}{\mathbb{P}}
\newcommand{\X}{\mathcal{X}}
\newcommand{\Lc}{\mathcal{L}}
\newcommand{\intX}{\int_{\X}}
\newcommand{\hti}{h}
\newcommand{\x}{\tilde{x}}
\newcommand{\Xt}{\tilde{X}}
\newcommand{\Qdt}{Q_{\Dt}}
\newcommand{\Id}{\mathrm{Id}}
\renewcommand{\H}{\mathcal{H}}
\newcommand{\e}{\mathrm{e}}
\newcommand{\lambdadt}{\lambda_{\Dt}}
\newcommand{\Nit}{N_{\mathrm{iter}}}

\newtheorem{assumption}{Assumption}

\newtheoremstyle{myplain}
{5pt}			
{5pt}			
{\normalsize}	
{}			
{\bfseries}		
{:}			
{.5em}		
{\thmname{#1}\thmnumber{ #2}\thmnote{~{(#3)}}}

\begin{document}

\title{Adaptive sampling of large deviations}

\author{Gr\'egoire Ferr\'e}
\email{gregoire.ferre@enpc.fr}
\affiliation{CERMICS, \'Ecole des Ponts ParisTech, Universit\'e Paris-Est, F-77455 Marne-la-Vall\'ee, France}

\author{Hugo Touchette}
\email{htouchette@sun.ac.za, htouchet@alum.mit.edu}
\affiliation{National Institute for Theoretical Physics (NITheP), Stellenbosch 7600, South Africa}
\affiliation{Institute of Theoretical Physics and Department of Physics, University of Stellenbosch, Stellenbosch 7600, South Africa}

\date{\today}

\begin{abstract}
We introduce and test an algorithm that adaptively estimates large deviation functions characterizing the fluctuations of additive functionals of Markov processes in the long-time limit. These functions play an important role for predicting the probability and pathways of rare events in stochastic processes, as well as for understanding the physics of nonequilibrium systems driven in steady states by external forces and reservoirs. The algorithm uses methods from risk-sensitive and feedback control to estimate from a single trajectory a new process, called the driven process, known to be efficient for importance sampling. Its advantages compared to other simulation techniques, such as splitting or cloning, are discussed and illustrated with simple equilibrium and nonequilibrium diffusion models.
\end{abstract}

\maketitle

\section{Introduction}

We consider in this paper the problem of estimating large deviation functions characterizing the fluctuations of time-integrated functionals of Markov processes in the long-time limit. These functions have a wide range of applications in engineering and physical sciences, where they are used to predict the probability of rare events \cite{shwartz1995,dembo1998,hollander2000} and to understand how these events arise via transition paths or modified processes \cite{jack2010b,chetrite2013,chetrite2014}. Large deviation theory also underlies now much of the research on nonequilibrium systems driven in steady states by non-conservative forces or boundary reservoirs \cite{touchette2009,derrida2007,harris2013}. In this context, large deviation functions play the role of nonequilibrium potentials, similar to the entropy or free energy, that characterize the steady state and fluctuations of physical quantities, such as energy or particle currents exchanged with reservoirs, as well as the presence of symmetries \cite{harris2007} and phase transitions in fluctuations \cite{garrahan2007,hedges2009,espigares2013,aminov2014}.

Recently, many efforts have been devoted to developing numerical algorithms for estimating large 
deviation functions that go beyond the direct sampling of probabilities, which require 
prohibitively large samples. The most popular algorithms are variations of two basic 
approaches used in rare event simulations, namely: 1) splitting \cite{dean2009,cerou2007,aristoff2015} or 
cloning \cite{grassberger2002,giardina2006,lecomte2007a} algorithms, which use 
population dynamics to estimate probabilities or generating functions that have a multiplicative structure in time, and 2) importance sampling \cite{bucklew2004,juneja2006,asmussen2007} (including transition path sampling \cite{bolhuis2002}), which modifies the process to be simulated, so as to transform rare events into typical events that can be simulated 
efficiently. Deterministic methods not based a priori on sampling can also be used, including spatial discretizations of 
various spectral and optimal control representations of large deviation functions, which 
work well for low-dimensional systems, in addition to action minimization methods, which can 
be applied in the low-noise or low-temperature limit \cite{heymann2008,eijnden2012,grafke2015}.

In this paper, we propose an algorithm that combines spectral methods with importance sampling to 
efficiently estimate large deviation functions in an adaptive way. The core of the algorithm comes 
from recent works on 
learning algorithms for risk-sensitive control of Markov chains \cite{borkar2004,ahamed2006,basu2008,borkar2010}, which we adapt to 
continuous-time diffusion processes and to the problem of estimating large deviation functions. The algorithm works by estimating or learning ``on the fly'' 
a modified process, called the auxiliary or \emph{driven process}, which corresponds to the 
process that is asymptotically equivalent to the original process conditioned on the rare event 
of interest \cite{chetrite2014} or, alternatively, to the exponential tilting of the original process, known to be 
efficient for importance sampling \cite{chetrite2015}. This modified 
process is given by a principal eigenvalue problem related to the Feynman--Kac 
equation or, equivalently, by a stochastic optimal control 
problem \cite{chetrite2015} that we solve iteratively using stochastic approximation and feedback control methods.

The main advantage of this algorithm, compared to splitting or
cloning, is that it does not require the simulation of many copies 
of the process considered -- it runs on one long trajectory of that process, modified
with a feedback-reinforcement rule, to 
adaptively learn the driven process, thereby 
reducing significantly the complexity of estimating large deviation functions. The calculation 
of error bars for the estimated quantities is also simplified compared to other techniques, as 
the algorithm is based on simple time averages and stochastic 
approximations \cite{chauveau2003estimation,roberts2004general,benveniste2012adaptive}.
Finally, the errors incurred by discretizing continuous-time processes and functionals can be analysed
in a precise way, in principle, via Feynman--Kac semi-groups \cite{ferre2017}.

We discuss these advantages and test the algorithm in Sec.~\ref{sec:numerics} with simple equilibrium and nonequilibrium
diffusions, after introducing the general model and notations 
in Sec.~\ref{sec:setting} and the algorithm in Sec.~\ref{sec:adaptive}. The results at
this point are preliminary and are presented as a proof of concept of the algorithm. More
detailed results about the time-discretization and sampling errors will
be addressed in future works, together with more complex 
applications involving interacting particle systems and higher-dimensional diffusions.

\section{Framework}
\label{sec:setting}

\subsection{Model and notations}
We consider an ergodic diffusion $(X_t)_{t\geq 0}$ evolving in a state space $\X\subset \R^d$ according to the following stochastic differential equation (SDE):
\be
\label{eq:SDE}
dX_t = b(X_t)\,dt + \sigma\, d W_t,
\ee
where $b: \X \to \R^d$ is a smooth function, called the \emph{drift}, $W_t$ is an $m$-dimensional Brownian motion, and $\sigma$ is a $d\times m$ matrix, assumed to be constant for simplicity.\footnote{See \cite{chetrite2014} for a treatment of diffusions with multiplicative noise.} The generator of this diffusion reads
\be
L = b \cdot \nabla + \frac{1}{2}\nabla\cdot D \nabla,
\ee
where $\cdot$ denotes the scalar product and $D=\sigma\sigma^\transp$, with $\transp$ as the transpose, is the diffusion matrix, assumed to be positive definite. This is the generator of the 
evolution semi-group $P_T$, defined by
\be
\label{eq:Pt}
P_T\varphi(x) = \E \left[ \varphi ( X_T )\, \middle|\,  X_0 = x \right],
\ee 
for all time $T\geq 0$ and any smooth test function $\varphi$. The dual $L^\dag$ of $L$ in the space $L^2(dx)$ of square-integrable functions with respect to the Lebesgue measure is the generator of the Fokker-Planck equation
\be
\p_t \rho(x,t) =L^\dag \rho(x,t),
\ee
which gives the evolution of the probability density $\rho(\cdot,t)$ of $X_t$ starting from some 
initial density $\rho(\cdot,0)$ for $X_0$.

Our goal here is to study the fluctuations of time-integrated functionals of $X_t$, called \emph{observables}, having the general form
\be
\label{eq:AT}
A_T =\frac{1}{T}\int_0^T f(X_t)\,dt + \frac{1}{T}\int_0^T g(X_t) \circ dX_t,
\ee
where $f:\X\to\R$ and $g:\X\to\R^d$ are reasonably smooth functions (\textit{e.g.}\ continuous) 
and $\circ$ denotes the Stratonovich product \cite{pavliotis2014}. 
Such a functional defined over the time horizon $[0,T]$ can represent, for example, a control 
cost associated with the state $X_t$ and its increments \cite{chernyak2014} or a physical quantity integrated in time, 
such as the work performed on a particle by external forces or the heat exchanged by a particle with its environment \cite{sekimoto2010}. 

Assuming that the process is ergodic with respect to an invariant measure $\mu(dx)=\rho^*(x)\,dx$
with smooth density $\rho^*$, we have
\be
A_T\xrightarrow{T\to \infty} \intX f(x)\rho^*(x)\,dx+ \intX g(x)\cdot J^*(x)\,dx= a^*,
\ee
almost surely, where 
\be
J^*(x) = b(x) \rho^*(x)-\frac{D}{2}\nabla \rho^*(x)
\ee
is the stationary current field associated with $\rho^*$ \cite{chetrite2015}. The theory of large deviations
\cite{dembo1998} refines this ergodic theorem, generalized
here with the additional $g$ term, by providing estimates for the rate at which the probability
distribution of $A_T$ concentrates on its ergodic value $a^*$. Such estimates can be derived
under general conditions (see~\cite{dembo1998}) and take, in the simplest case, the form
\be
\lim_{T\ra\infty} -\frac{1}{T}\log \proba(A_T\in B)=\min_{a\in B}\, I(a),
\label{eqldp1}
\ee
for any Borel subset $B$ of $\R$, where $I:\R\ra\R_+$ is a positive function such that $I(a^*)=0$. When this limit exists, $A_T$ is said to satisfy the \emph{large deviation principle} (LDP) with rate function $I$. Formally, this means that 
\be
\label{eq:ldp1}
\proba(A_T\in da) = \e^{-TI(a)+o(T)}da,
\ee
where $o(T)$ denotes corrections in the exponential that grow slower than linearly
in $T$. Thus, we see that the rate function provides useful information about the fluctuations of $A_T$: the likelihood that $A_T=a$ decays exponentially with time for all $a\neq a^*$, since $I(a)>0$ in this case, and converges otherwise to 1 as $T\ra\infty$, since $I(a^*)=0$. Moreover, the rate function is in general not a parabola, meaning that it describes fluctuations that are generally not Gaussian. In this sense, large deviation theory is often seen as an extension of both the ergodic theorem, which describes the concentration of $A_T$ towards its mean, and the central limit theorem, which describes the local Gaussian fluctuations of $A_T$ around its mean \cite{dembo1998}.

\subsection{Large deviation functions and driven process}

In practice, the rate function $I(a)$ can be calculated in many different ways other than by direct 
sampling, which requires exponentially large samples with $T$ \cite{bucklew2004}. The most common method proceeds from the
\emph{scaled cumulant generating function} (SCGF), defined for $k\in\R$ by
\be
\label{eq:scgf}
\lambda(k)=\underset{T\ra\infty}{\lim}\ \frac{1}{T}\log\E\left[ \e^{k T A_T} \right].
\ee 
By the G\"artner-Ellis Theorem \cite{dembo1998}, the
Legendre--Fenchel transform of this function yields the rate function:
\be
\label{eq:lf1}
I(a) = \sup_{k\in\R}\, \{ka-\lambda(k)\},
\ee
provided, essentially, that $\lambda(k)$ is differentiable; see~\cite{dembo1998} for more precise conditions. 

For the SDE \eqref{eq:SDE} and the additive
functional \eqref{eq:AT}, the SCGF is known to be given by the principal eigenvalue of the operator
\be
\label{eq:tilted}
\Lc_k =  b  \cdot (\nabla+kg) + \frac{1}{2} (\nabla + k g)\cdot D (\nabla + k g) + k f,
\ee
which is the generator of the \emph{Feynman--Kac semi-group} $P_T^k$, defined by
\be
\label{eq:Ptk}
P_T^k\varphi(x)=\E \left[ \varphi(X_T)\,\e^{T kA_T}
\, \middle| \, X_0 =x \right],
\ee
for any smooth function $\varphi$; see Appendix A.2 of \cite{chetrite2014}. In the end, the
rate function can thus be computed by solving the spectral problem
\be
\label{eq:spectral}
\Lc_k h_k = \lambda(k) h_k,
\ee
where $\lambda(k)$ is the principal eigenvalue of $\Lc_k$ and $h_k$ its corresponding eigenvector. This holds
provided that this operator has reasonable spectral properties, made precise in the following
assumption.

\begin{assumption}
\label{as:spectral}
The operator $\Lc_k$ defined in \eqref{eq:tilted} acting on $L^2(\mu)$ has an isolated largest
eigenvalue $\lambda(k)$. Its multiplicity is one and it is associated with a regular eigenvector 
$h_k\in L^2(\mu)$ such that for all $x\in\X$, $h_k(x) >0$.
\end{assumption}


This assumption holds for many systems, in particular when $\X$ is bounded \cite{bierkens2013,ferre2017}
or when $b$ and $g$ are gradient fields with appropriate growth conditions; see \cite[Sec.~2.5]{lelievre2016}. In practice, the spectral
problem \eqref{eq:spectral} can be solved numerically using standard projection or discretization (Galerkin)
methods, which work well for low-dimensional systems \cite{chatelin2011spectral}, or more involved
real space renormalization methods when dealing with higher-dimensional 
systems \cite{gorissen2011}. Note that, for $g=0$, $\Lc_k$ is 
the usual Feynman--Kac generator with source term $kf$. Moreover, for $k=0$, $\Lc_{0}=L$ is
the generator of the SDE, so that $\lambda(0)=0$ and $h_0=1$, the constant unit function.

The numerical method that we propose in the next section attempts to estimate the spectral 
elements $\lambda(k)$ and $h_k$ in a different way using the fact that they are solutions of 
the family of eigenproblems
\be
\label{eq:eigenFK}
P_T^k h_k = \e^{T\lambda(k)}h_k,\qquad \forall\, T>0,
\ee
which can be approximated stochastically. The method also exploits a connection between large 
deviations and control theory showing that $\lambda(k)$ is the ergodic limit of an optimal 
control cost satisfying a stochastic Hamilton--Jacobi--Bellman 
equation \cite{fleming2006}, and that $h_k$ 
determines the controlled diffusion achieving the optimal cost, which is the driven process 
mentioned in the introduction. These results are explained in detail in \cite{chetrite2015} (see also references therein); 
for the purpose of this paper, we only state them without proofs. 

The controlled diffusion, denoted by $(\Xt_t)_{t\geq 0}$, satisfies the SDE
\be
\label{eq:driven}
d\Xt_t = b_k(\Xt_t)\, dt + \sigma\, dW_t,
\ee
where
\be
\label{eq:bk}
b_k(x) = b(x) +D[kg(x)+\nabla\log h_k(x)]
\ee
is the optimal control drift defined for all $x\in\X$. Under
Assumption \ref{as:spectral}, this modified diffusion is ergodic with respect to a new invariant measure $\mu_k$, whose density is
\be
\rho_k(x)=h_k(x) l_k(x),
\ee
where $l_k$ is the dual of $h_k$ in $L^2(dx)$ satisfying $\Lc_k^\dag l_k=\lambda(k) l_k$ \cite[Sec.~3.3]{chetrite2014}. These two functions are normalized such that
\be
\intX h_k(x)l_k(x) \, dx=1,\qquad \intX l_k(x)\, dx =1.
\ee

Because of the change of process, the observable $A_T$ must concentrate on a new value, which can be shown to be given by $\lambda'(k)$, that is,
\be
\label{eq:ATtilde}
A_T\xrightarrow{T\to \infty} \lambda'(k)= a_k,
\ee
almost surely with respect to $(\Xt_t)_{t\geq 0}$. Similarly, the control cost
\be
\label{eq:cost1}
C_T = kA_T -R_T,
\ee
where
\be
\label{eq:cost2}
R_T  =\frac{1}{2T}\int_0^T [b(\Xt_t)-b_k(\Xt_t)]\cdot D^{-1}[b(\Xt_t)-b_k(\Xt_t)]\, dt,
\ee
reaches in the ergodic limit the value $\lambda(k)$, so that
\be
C_T\xrightarrow{T\to \infty} \lambda(k)
\label{eqerg2}
\ee
almost surely under $(\Xt_t)_{t\geq 0}$ \cite[Sec.~4]{chetrite2015}. Finally, it can be shown by Legendre duality that the rate function at the value $a_k=\lambda'(k)$ is given by the ergodic limit of $R_T$ above, leading to
\be
\label{eq:rfct1}
R_T \xrightarrow{T\to \infty} I(a_k),
\ee
almost surely with respect to $(\Xt_t)_{t\geq 0}$ \cite[Sec.~4]{chetrite2015}. Note that the diffusion is not modified for $k=0$, so that $a_0=a^*$, $\lambda'(0)=a^*$, and $I(a^*)=0$.

These ergodic limits provide direct estimators of the SCGF and the rate function, based on a single trajectory of the driven process, which can be simulated for different values of the parameter $k\in\R$. For the SCGF, there are in fact three possible estimators:
\begin{enumerate}
\item From \eqref{eq:ATtilde}: The value of $A_T$, integrated numerically with the condition $\lambda(0)=0$.
\item From \eqref{eqerg2}: The value of $C_T$.
\item The eigenvalue returned by the algorithm proposed in Sec.~\ref{sec:adaptive}.
\end{enumerate}
In practice, we find that the last estimator is more stable, although the first and second are more adapted to obtain error bars. 

For the rate function, we have two possible estimators: 
\begin{enumerate}
\item The Legendre transform \eqref{eq:lf1} of the SCGF, given in parametric form by
\be
\label{eq:legendre}
I(a_k) = ka_k -\lambda(k),
\ee
where $a_k$ is either estimated from $A_T$ or by taking the numerical derivative of $\lambda(k)$.
\item From \eqref{eq:rfct1}: The value of $R_T$ obtained for the value of $A_T$, giving the couple $(A_T,R_T)$.
\end{enumerate} 
In practice, we find that the first estimator based on the Legendre transform is more reliable.
By comparison, the computation of $I(a)$ based on \eqref{eq:cost2} involves
the optimal drift $b_k$ and, therefore, the logarithmic derivative of $h_k$, which is more difficult to estimate in a stable way.

In all cases, error bars can be constructed from the same trajectory by estimating, in principle,
the variance of $A_T$, $C_T$ and $R_T$ using covariance techniques for Markov
processes \cite{chauveau2003estimation,roberts2004general,benveniste2012adaptive}.
This is an advantage over splitting and cloning algorithms, for which the calculation of errors
bars is difficult, as they involve correlated copies or ``clones'' of the simulated
process \cite{rousset2006control,nemoto2016,nemoto2017}.

In closing, it is interesting to note that the driven process can also be interpreted as the change of process in the importance sampling of the probability $\proba (A_T\in B)$ that is optimal in the sense of logarithmic or asymptotic efficiency \cite{bucklew2004}. Therefore, it can be used not only to estimate the SCGF and rate function, but also to estimate the actual probability $\proba (A_T\in B)$ in an efficient way \cite{borkar2004}. The optimal change of process in this case is known to correspond to the exponential tilting of the original process \cite{bucklew2004}, which is a time-dependent process in general; see Appendix D of \cite{chetrite2014}. In the ergodic limit, this process converges to a homogeneous process corresponding exactly to the driven process \eqref{eq:driven}. We refer to \cite{chetrite2015} for more details about these results.

\section{Adaptive algorithm}
\label{sec:adaptive}

We are now ready to present the algorithm for estimating the SCGF and the rate 
function of $A_T$ for continuous-time diffusions. The algorithm is based, as mentioned, on
prior algorithms proposed in \cite{borkar2004,ahamed2006,basu2008,borkar2010} for Markov chains
and exploits the fact that
\be
\label{eq:ergfk1}
\e^{-T \lambda(k)} (P_T^k\varphi) \xrightarrow{T\to\infty} h_k \intX \varphi(x)\, l_k(x)\, dx,
\ee
for any smooth test function $\varphi$. This result follows under Assumption \ref{as:spectral}; see \cite[Sec.~III.B]{chetrite2014}
or \cite[Sec.~6.1]{ferre2017} for a proof. The algorithm that we propose works from this limit by approximating the action of the Feynman--Kac semi-group $P_T^k$ in a stochastic way as a time average computed over a long trajectory of the diffusion. Moreover, it continuously modifies the diffusion as we estimate $h_k$ to construct the driven process $(\Xt_t)_{t\geq 0}$, which underlies the estimators of the large deviation functions.

The algorithm is presented next. Our contribution compared 
to \cite{borkar2004,ahamed2006,basu2008,borkar2010} is to consider general time-continuous processes,
time-additive functionals of these processes that depend on both their state and increments, and,
more importantly, to construct the driven process explicitly so as to estimate the SCGF and the rate function adaptively using the estimators introduced in the previous section.

\subsection{Time discretization}
\label{sec:discr}

The first step required in the algorithm is to discretize in time the SDE \eqref{eq:SDE} and its 
associated Feynman--Kac semi-group \eqref{eq:Ptk} by transforming the Markov diffusion 
$(X_t)_{t\geq 0}$ into a Markov chain $(x_n)_{n\in\N}$ with a small time step. Many 
discretization schemes can be used for the SDE; see, for instance, \cite{kloeden1992}. 
Here we use the standard Euler--Maruyama scheme with constant time step $\Delta t$, given by
\be
\label{eq:EM}
x_{n+1}=x_n + b(x_n)\Dt + \sigma \sqrt{\Dt}\,\xi_n,
\ee
where $(\xi_n)_{n\geq 0}$ is a sequence of independent standard  $d$-dimensional Gaussian
random variables. The corresponding discretization of the evolution semi-group $P_T$ over a
time step $\Dt$ is denoted by $\Qdt$, so that
\be
\label{eq:Q}
\Qdt\varphi(x)= \E\left[ \varphi(x_{n+1}) \middle|x_n = x \right],
\ee
for any test function $\varphi$ and $x\in\X$. We refer to \cite{kloeden1992,lelievre2016} for more
information about the discretization of SDEs and their weak error analysis.

Many discretizations also exist for the Feynman--Kac semi-group $P_T^k$. Here, we use the 
natural scheme where the diffusion is discretized as above and the integral of $A_T$ is 
discretized as a Riemann sum with the left-point rule for the integral involving $f$ and the 
mid-point rule for the Stratonovich integral involving $g$ \cite{ferre2017}.
The action of $P_T^k$ is thus replaced by 
\be
\label{eq:Qk}
\Qdt^k\varphi(x)= \E \left[ \e^{k \left[f(x_n)\Dt + g\left( \frac{x_{n+1}+x_n}{2}\right)\cdot
(x_{n+1}-x_n) \right]}
\varphi(x_{n+1})\ \middle|\ x_n = x 
\right].
\ee
For our purposes, $h_k$ will be approximated by recursive applications
of $\Qdt^k$, based on the following assumption. 

\begin{assumption}
\label{as:spectralapprox}
There exist a time step $\Dt^* >0$ and $p >0$ such that, for $0< \Dt \leq \Dt^*$,
\be
\label{eq:approxspectral}
\Qdt^k h_{k,\Dt} = \e^{\Dt \lambdadt(k)} h_{k,\Dt},
\ee
where
\be
\label{eq:biaseigen}
h_{k,\Dt}= h_k + O(\Dt^p), \quad \lambdadt(k) = \lambda(k) + O(\Dt^p).
\ee
\end{assumption}

This assumption means that the time-discretized operator $\Qdt^k$ admits $h_k$ as an 
approximate eigenvector with approximate eigenvalue $\lambda(k)$. This applies, for example, when $\X$ is compact. In this case, precise estimates for the errors in \eqref{eq:biaseigen} are obtained for $g=0$ in \cite{ferre2017} and can be extended to $g\neq 0$.

In the following, we will drop the subscript $\Delta t$ on $h_{k,\Delta t}$ and $\lambda_{\Delta t}(k)$ to simplify the notations, and will present the algorithm essentially as if $h_k$ were an exact eigenvector
of $\Qdt^k$ with exact eigenvalue $\lambda(k)$. However, we should keep in mind that this is only approximately true due to the errors in $\Dt$. We will comment on this in Sec.~\ref{sec:numerics} with specific numerical examples.

\subsection{Stochastic approximation and annealing}

The main ingredient of the algorithm is the limit \eqref{eq:ergfk1} of the Feynman--Kac
semi-group, which shows that $h_k$ and $\lambda(k)$ can be computed by successively
applying $Q^k_{\Dt}$ to an initial guess $\varphi$, so as to obtain
\be
\label{eq:PM}
(\Qdt^k)^n\varphi\sim \e^{n\Dt \lambda(k)} h_k \intX \varphi(x) \, l_k(x)\, dx
\ee
as $n\to\infty$. To perform this iteration, which is a functional version of the well-known
\emph{power method} for matrices \cite{demmel1997applied}, we apply a stochastic
approximation \cite{polyak1992acceleration,benveniste2012adaptive,basu2008} whereby the
expectation appearing in the action of $\Qdt^k$ is replaced by the iterates of the Markov chain:
\be
\label{eq:approxsto}
\begin{aligned}
\Qdt^k \varphi (x_n) & = \E \left[ \e^{k \left[f(x_n)\Dt + g\left( \frac{x_{n+1}+x_n}{2}\right)\cdot
(x_{n+1}-x_n) \right]}
\varphi(x_{n+1})\ \middle|\ x_n \right] \\ & \approx \e^{k \left[f(x_n)\Dt + 
g\left( \frac{x_{n+1}+x_n}{2}\right)\cdot (x_{n+1}-x_n) \right]}
\varphi(x_{n+1}),
\end{aligned}
\ee
where $x_{n+1}$ is a random variable distributed according to $\Qdt(x_n, \cdot)$. This approximation is known to reproduce the 
expectation as a statistical average in the ergodic limit $n\ra\infty$ \cite{ahamed2006,basu2008,benveniste2012adaptive}. 

In our case, we simulate not the Markov
chain $x_n$, but a modified chain, corresponding to the discretization of the driven process \eqref{eq:driven},
which we express as 
\be
\label{eq:biaseddyn}
\x_{n+1}= \x_n + [b(\x_n) + F_n(\x_n)]\Dt + \sigma\sqrt{\Dt}\,\xi_n,
\ee
where
\be
F_n = D(kg + \nabla \log h_k^n)
\label{eqbias1}
\ee 
is the extra biasing force derived, according to (\ref{eq:bk}), from the estimate $h_k^n$ of $h_k$ at time $n$. In this
case, the evolution \eqref{eq:approxsto} is modified by Girsanov
formula \cite{pavliotis2014} to
\be
\label{eq:approxstobiased}
\begin{aligned}
\Qdt^k\varphi(\x_n) & = \E \left[ \e^{k \left[f(\x_n)\Dt + g\left( \frac{\x_{n+1}+\x_n}{2}\right)\cdot
(\x_{n+1}-\x_n) \right]}
\varphi(\x_{n+1}) R_n(\x_n, \x_{n+1}) \ \middle|\ \x_n = x 
\right] \\
&\approx  \e^{k \left[f(\x_n)\Dt + 
g\left( \frac{\x_{n+1}+\x_n}{2}\right)\cdot (\x_{n+1}-\x_n) \right]}
\varphi(\x_{n+1}) R_n(\x_n, \x_{n+1}),
\end{aligned}
\ee
where
\be
\label{eq:Ln}
R_n(\x_n, \x_{n+1})= \exp\left( -\frac{1}{2\sigma^2}F_n^2(\x_n) \Dt - \sqrt{\frac{\Dt}{\sigma^2}} F_n(\x_n)\cdot\xi_n \right)
\ee
is the Radon--Nikodym derivative of the transition kernel of $x_n$ with respect to that of $\x_n$.

In the end, we also apply an annealing scheme, commonly used in stochastic approximations, which consists in replacing the update rule \eqref{eq:PM}, defined by $\varphi^0=\varphi$ and
$\varphi^{n+1}= \Qdt^k\varphi^n$, by the scheme
\be
\varphi^{n+1} = \varphi^n + a_n\left( \Qdt^k\varphi^n - \varphi^n \right),
\ee
where $(a_n)_{n\in\N}$ is a decreasing sequence, often called the adaptation or learning sequence, which acts as a smoothing parameter, filtering here the noisy update of the eigenfunction.
This sequence is usually chosen in such a way that
\be
\label{eq:ancond}
\sum_{n\geq 0} a_n = \infty,\qquad \sum_{n\geq 0} a_n^2 < \infty,
\ee
with the understanding that $a_n$ should not be decreased too slowly, so as to limit noise, nor too fast, so as to reach the ``correct'' fixed point. These conditions can be relaxed under stability
assumptions or by an averaging procedure \cite{polyak1992acceleration}; see \cite{benveniste2012adaptive} for more details. 

\subsection{Spatial projection}
\label{sec:spectral}

The iteration just described for approximating $h_k$ can be performed numerically by discretizing the state space $\X$ into
small cells (grid discretization). For 
high-dimensional systems, however, it is more convenient to use a 
Galerkin-type approximation of the eigenvalue and eigenfunction \cite{basu2008}, obtained by
projecting the problem onto a set of basis functions $\{\phi_j\}_{j=1}^M$ with $\phi_j:\X\to\R$ \cite{chatelin2011spectral}. Let us denote by $\H_M=\mathrm{Span}\{\phi_j\}_{j=1}^M$ the space
spanned by these functions. Then the $M$-dimensional eigenproblem that we need to solve is
\be
\label{eq:discreigen}
\Qdt^k \hti = \e^{\Dt \lambda} \hti, \quad \hti \in \H_M,
\ee
using, for notational convenience, the same symbols $\hti$ and $\lambda$ for the exact and the projected spectral
elements. We also drop from now on the parameter $k$, which will be implicit.

The eigenfunction $\hti$ is expressed in that basis as
\be
\hti(x) = \sum_{j=1}^M \alpha_j \phi_j(x)=\alpha^\transp\phi(x),
\ee
where $\alpha= [\alpha_1, \hdots, \alpha_M]^\transp$ and
$\phi(x)=[ \phi_1(x),\hdots,\phi_M(x)]^\transp$. Multiplying \eqref{eq:discreigen} by $\phi_i$ for
$i\in\{1,\hdots, M\}$ and integrating over any measure $\eta$ on $\X$ yields
\be
\sum_{j=1}^M \alpha_j \intX \phi_i (\Qdt^k\phi_j) \,d\eta
=  \e^{\Dt \lambda} \sum_{j=1}^M \alpha_j \intX \phi_i \phi_j\, d\eta, \quad
i\in\{1,\hdots, M\}.
\ee
As a result, we see that the vector of coefficients $\alpha\in\R^M$ is the principal solution of the eigenproblem
\be
\label{eq:eigenproblem}
 A \alpha = \Lambda B \alpha,\qquad \lambda = \frac{1}{\Dt}\log \Lambda,
\ee
where
\be
\label{eq:defmatrices}
A= \intX \phi (\Qdt^k\phi^\transp)\, d\eta, \qquad B = \intX \phi\, \phi^\transp\, d\eta.
\ee
Note that the matrix $B$ is invertible as soon as the $\{\phi_j\}_{j=1}^M$ form a linearly
independent family. 

\subsection{Algorithm}
\label{sec:algo}

We are now ready to describe all the steps of the algorithm that estimates the principal eigenvalue of $\Lc_k$ and its corresponding eigenfunction. For a fixed $k\in\R$, we initiate the process at a position $x_0\in\X$ and define a first approximation $\hti^0$ of $h$ as
\be
\hti^0= (\alpha^0)^\transp \phi,
\ee
where $\phi$ is the vector of basis functions and $\alpha^0$ is the initial vector of coefficients, chosen such that $h^0$ is constant.  At each iteration, we then perform the following steps:

\begin{enumerate}
\item Draw a new position $\x_{n+1}$ according to the Markov chain \eqref{eq:biaseddyn}.

\item Compute the extra bias $F_n$ according to \eqref{eqbias1}, which in the function basis takes the form
\be
\label{eq:Fnloc}
F_n = D \left( kg + \frac{\sum_{j=1}^M \alpha_j^n \nabla \phi_j}{\sum_{j=1}^M \alpha_j^n  \phi_j}\right). 
\ee

\item Compute the Girsanov weight $R_n$ according to \eqref{eq:Ln}.

\item Compute the matrices $A_{n+1}$ and $B_{n+1}$ using the formulae
\begin{eqnarray}
A_{n+1} & =& \frac{1}{n+1} \sum_{m=0}^n \e^{k \left[f(\x_m)\Dt + 
        g\left( \frac{\x_{m+1}+\x_m}{2}\right) (\x_{m+1}-\x_m) \right]}
    \phi(\x_m) \phi(\x_{m+1})^\transp R_{m}(\x_m,\x_{m+1}),\nonumber\\  
B_{n+1} & =& \frac{1}{n+1}\sum_{m=0}^n \phi(\x_m) \phi(\x_{m})^\transp,
\label{eq:biasedBn}
\end{eqnarray}
which follow by projecting \eqref{eq:approxstobiased} and \eqref{eq:defmatrices}, respectively.

\item Update the coefficient vector $\alpha^n$, giving the decomposition of the iterate $h^n$, as
\be
\alpha^{n+1} = \alpha^n + a_n \left(
\frac{B_n^{-1} A_n}{\hti^n(x_0)} - \Id 
\right) \alpha^n,
\ee
where $\Id$ is the identity matrix of size $M$ \cite{basu2008}. 

\item Estimate the eigenvalue as
\be
\label{eq:lambdaestimate}
\lambda^{n+1}= \frac{n}{n+1}\lambda^n +
\frac{1}{(n+1)\Dt}  \log \left( \hti^{n}(\x_0)\right),
\ee
which follows from \eqref{eq:eigenproblem}.
\end{enumerate}

Repeating these steps, it can be proved that the iterates $h^n$ and $\lambda^n$ converge to the solution of 
the spectral problem \eqref{eq:discreigen}, following the analysis found 
in \cite{borkar2004,ahamed2006,basu2008}; see also \cite{benveniste2012adaptive}. Moreover, because 
$h^n\ra h$, the Markov chain $(\x_n)_{n\in\N}$ samples in the long run the Euler--Maruyama 
discretization of the driven process \eqref{eq:driven}.

\subsection{Remarks}

The following are technical remarks worth noting about the algorithm:

\begin{enumerate}
\item The matrix $A_n$ can be updated at each step using
\be
A_{n+1}= \frac{n}{n+1}A_n + \frac{1}{n+1} \e^{k \left[f(\x_n)\Dt + 
g\left( \frac{\x_{n+1}+\x_n}{2}\right) (\x_{n+1}-\x_n) \right]} \phi(\x_n) \phi(\x_{n+1})^\transp R_{n},
\ee
instead of the sum shown in (\ref{eq:biasedBn}). Similarly, the matrix $B_n$ can be updated, following  \cite{basu2008}, using the Shermann--Morrisson--Woodburry 
formula, which leads here to
\be
B_{n+1}^{-1} = \frac{n+1}{n} B_n^{-1} - \frac{n+1}{n}
 \frac{ B_n^{-1} \phi(\x_{n})\phi(\x_{n})^\transp B_n^{-1}}
{n + \phi(\x_{n})^\transp B_n^{-1} \phi(\x_{n})}.
\label{eqsmw}
\ee
The advantage of this formula is that the computation required for updating the
coefficient $\alpha^n$ scales with the number $M$ of basis vectors as $M^2$, as is common in the power method \cite{demmel1997applied},
whereas the typical cost of inverting the matrix $B$ scales as $M^3$.

\item We normalize the eigenfunction at every iteration by setting $\hti(\x_0)=1$ at an arbitrary location, taken here to be the initial state $\x_0$ \cite{borkar2004}. This prevents the norm of the eigenfunction from diverging or decaying to zero, as is common in the power method, and provides an estimate of $\lambda(k)$ through \eqref{eq:lambdaestimate}. This normalization step can be based on other norms, at the expense of computing integrals.

\item The algorithm can be run with the original (unbiased) process $X_t$, but the estimation of the matrices $A$ and $B$ in this case typically suffers from large statistical errors due to the high variance of the underlying estimator. This is a known problem related to the estimation of Feynman--Kac functionals and exponential integrals in general \cite{hartmann2012,hartmann2014,zhang2014,rohwer2014}. Biasing the dynamics with the driven process $\Xt_t$ reduces this variance in an optimal way (in the sense of asymptotic or logarithmic efficiency) by forcing the exploration of the process in important regions of the state space where the integrand of the generating function $\E[ \e^{k T A_T}]$ is largest \cite{chetrite2015}. 

\item As the dynamics is biased towards the driven process $\Xt_t$, the vector $\alpha^n$ of basis coefficients representing $h_k$ converges towards the solution of the eigenproblem \eqref{eq:eigenproblem}, with the matrices $A$ and $B$ computed as ergodic averages under $\Xt_t$, so that $\eta=\mu_k$ in \eqref{eq:defmatrices}. Even for a small number of basis functions, it would be a priori impossible to compute these matrices by numerical quadrature. This shows that the algorithm can be used to obtain good approximations of $h_k$ even for high-dimensional systems, provided that enough basis functions are used to represent the support of $\mu_k$, which is typically concentrated on a subset of $\X$.

\item The algorithm is stable despite the fact that it includes the Girsanov reweighting factor, which is exponential in time. The reason for this stability, already noted in \cite{ahamed2006}, is that the Girsanov weight is computed and accumulated incrementally over single time steps.

\item The learning sequence $(a_n)_{n\in\N}$ is chosen here in the following way:
\begin{enumerate}
\item For $0\leq n \leq N_1$, we take $a_n=0$, so there is no adaption at the level of $h$ and $\lambda$, although the matrices $A$ and $B$ are evolved. This ``burn-in'' period
allows for a better initial guess of the various functions estimated through their ergodic averages.
\item For $N_1 < n \leq N_1+\Nit-N_2$, we take $a_n=1$, that is, the full information of the
process is taken into account.
\item For $ N_1+\Nit-N_2 < n \leq N_1+\Nit$, we take $a_n= C/(n - (N_1 +\Nit- N_2))$ with 
constant $C>0$, so the process learns less with time, smoothing the noise in the long run. In the following, we choose $C=1$.
\end{enumerate}
The times $N_1$ and $N_2$ can be fixed or can be chosen dynamically according to some stopping rule. 

\item In practice, we can perform independent simulations with different values of $k$ to obtain an interpolation of the SCGF over some range, say, $[k_{\min},k_{\max}]$, which can then be used to obtain the rate function by Legendre transform. Alternatively, we can do a simulation in which $k$ is slowly increased from $k=0$ in a ``quasi-static'' way, so as to adaptively update the biasing force over a range of values for $k$ \cite{nemoto2014} or to reach some prescribed value of the SCGF for an unknown $k$ \cite{borkar2004}.
\end{enumerate}

\section{Applications}
\label{sec:numerics}

We apply in this section the algorithm to two simple test cases involving one-dimensional
diffusions. The first is the Ornstein-Uhlenbeck process, for which the large deviation functions of the area per unit time are known exactly \cite{chetrite2014}, while the second is a driven diffusion on the circle, often used in physics as a model of nonequilibrium systems, including Josephson junctions perturbed by thermal noise and Brownian particles controlled by external forces \cite{risken1996,reimann2002,ciliberto2010}. We discuss for both the convergence and efficiency of the algorithm.

\subsection{Ornstein-Uhlenbeck process}
\label{sec:OU}

The first example that we consider is the mean area or mean position 
\be
A_T = \frac{1}{T}\int_0^T X_t\, dt
\ee
of the Ornstein-Uhlenbeck process on $\R$ satisfying the SDE
\be
\label{eq:OU}
dX_t = - 2 \theta X_t\, dt + \sqrt{2}\,dW_t, 
\ee 
where $\theta>0$. In the notations of Sec.~\ref{sec:setting}, we thus have $b(x)=-2\theta x$, $\sigma=\sqrt{2}$, $f(x)=x$ and $g(x)=0$, so that
\be
L=-2\theta x\frac{d}{dx}+\frac{d^2}{dx^2},
\ee
and $\Lc_k=L + k x$.

For this process and observable, it can be checked \cite{chetrite2014} that the SCGF, corresponding to the dominant eigenvalue of $\Lc_k$, is
\be
\label{eq:analytical}
\lambda(k)=\frac{k^2}{4\theta^2},
\ee
so that $I(a) = \theta^2 a^2$ from \eqref{eq:legendre}. This is expected,
since the integral of a Gaussian process is also Gaussian. Moreover, the associated dominant
eigenfunction is
\be
h_k(x)=\e^{\frac{k}{2\theta} x},
\ee
leading with \eqref{eq:bk} to the optimal drift
\be
\label{eq:driftOU}
b_k(x)=-2\theta x +2(\log h_k(x))'=-2\theta x+  \frac{k}{\theta}.
\ee
This shows that a fluctuation of $A_T$ is created in an optimal way by adding a constant to the drift, which ``moves'' the Gaussian stationary density of the Ornstein-Uhlenbeck process
\be
\label{eq:rho}
\rho^*(x)=\sqrt{\frac{\theta}{\pi}}\,\e^{-\theta x^2}
\ee 
to 
\be
\label{eq:rhok}
\rho_k(x)=\sqrt{\frac{\theta}{\pi}}\,\e^{-\theta (x - m_k)^2}, \qquad m_k=\frac{k}{2\theta^2},
\ee
leading to
\be
A_T\xrightarrow{T\to \infty} m_k=a_k
\ee 
almost surely with respect to $\Xt_t$, in agreement with (\ref{eq:ATtilde}).

\begin{figure}[t]
\centering
\includegraphics[width=2.5in]{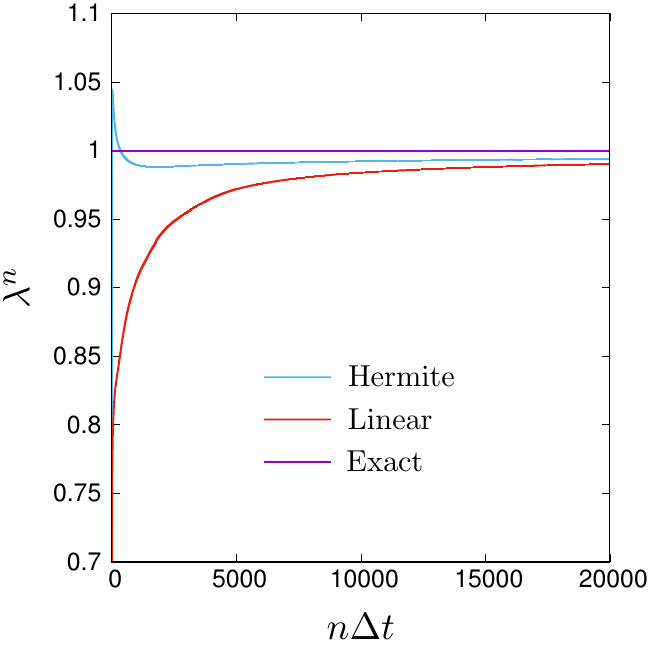}%
\hspace*{0.6in}%
\vspace{0pt}%
\includegraphics[width=2.5in]{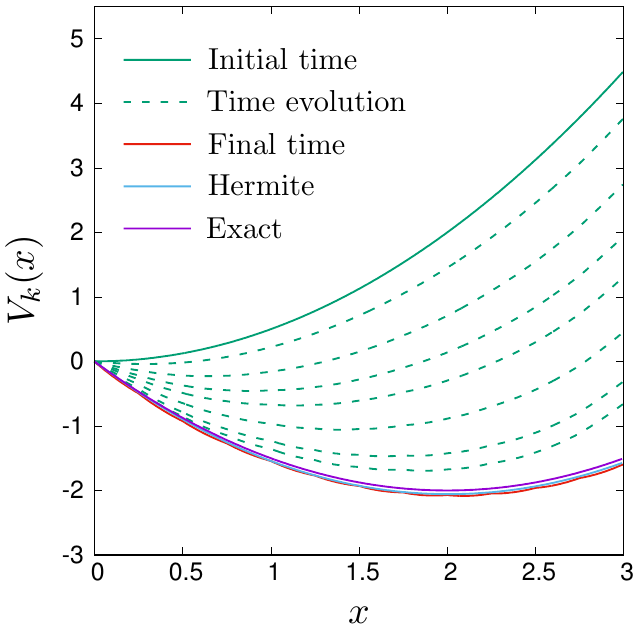}
\caption{Large deviations of the mean area for the Ornstein-Uhlenbeck process.
Left: Evolution of the estimated SCGF in time.
Right: Evolution of the biasing potential $V_k(x)$ in time for the linear basis (curves from top to bottom), compared with the final estimate for the Hermite basis and the exact result.}
\label{fig:biasedlearning}
\end{figure}

We compare the algorithm against these exact results using a simple mesh
discretization (first-order finite elements), defined by the basis functions
\be
\label{eq:hatfunc}
\phi_j(x)= \left\{
\begin{array}{lll}
 \frac{x - x_j}{\delta} +1, & & x\in [x_j -\delta, x_j], \\
 -\frac{x - x_j}{\delta} +1, & & x\in [x_j , x_j + \delta], \\
0, & & \mathrm{otherwise},
\end{array}
\right.
\ee
where the points $x_j$ define the centers of each ``cell'' of width $\delta>0$. In the following,
we refer to this basis simply as the \emph{linear basis}. To illustrate the flexibility of the
algorithm, we also perform simulations using a Hermite polynomial
basis, which forms a complete orthonormal 
basis in $L^2(\rho^*)$. We run the algorithm with 
$k=1$, $\theta=1/2$, $x_0=0$, $T=2\times 10^4$, $T_1=N_1\Delta t=T_2=N_2\Delta t=2\times 10^3$ with $\Dt=5\times 10^{-3}$, which a standard time step used in simulations relative to the basic timescale of the dynamics, corresponding here to $1/(2\theta)=1$. For the linear basis, we use $M=61$ equally spaced cells with $\delta=0.25$ around $x=0$, whereas for the
Hermite basis we use only $M=10$ basis functions.

Figure~\ref{fig:biasedlearning} illustrates the results of a typical simulation, starting on the left with the evolution of the estimate
of the SCGF, given by \eqref{eq:lambdaestimate}, as time increases. We observe a very good agreement in the long run with the exact value, which for the parameters used is equal to 1, with a faster convergence observed for the Hermite basis compared to the linear basis. As mentioned before, we can also recover
the SCGF by recording the stationary value of $A_T$, which corresponds as above to
$m_k=a_k=\lambda'(k)$, and numerically integrate the result in $k$ from $\lambda(0)=0$. The results obtained are similar to those obtained from the
eigenvalue estimate \eqref{eq:lambdaestimate}, and are not shown for this reason.

To understand the convergence of the eigenvalue at the process level, we show in the right plot of
Fig.~\ref{fig:biasedlearning} the evolution of the effective potential $V_k(x)$
associated with the modified drift $b_k(x)$ according to
\be
b_k(x)=-V_k'(x).
\ee
We show the results for both the linear
basis and the Hermite basis, with the zero of the potential arbitrarily set at $V_k(0)=0$.  We see from these that the tail of the potential takes longer to
be estimated correctly, as the process starts to explore values away from $m_0=0$. After
the convergence time, there is a good agreement with the parabola
\be
\label{eq:Vk}
V_k(x)= \theta (x - m_k)^2,
\ee
which is the exact solution predicted by \eqref{eq:rhok}. The small errors are due to the finite
time step $\Dt$ used and the finite basis function set. The small ``wiggles'' seen in the
potential obtained with the linear basis come from the fact that this basis is piecewise
linear, so that $V_k(x)$ is piecewise logarithmic. The potential obtained with the Hermite basis
is smoother, as expected.

\begin{figure}[t]
\centering
\includegraphics[width=2.5in]{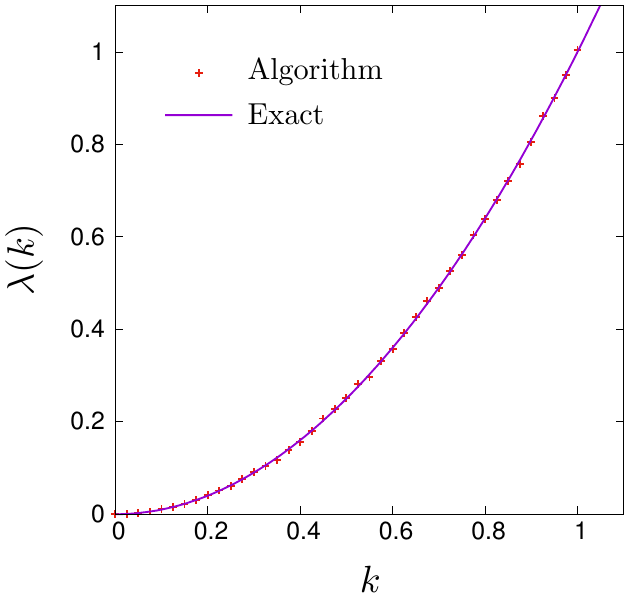}%
\hspace*{0.6in}%
\includegraphics[width=2.6in]{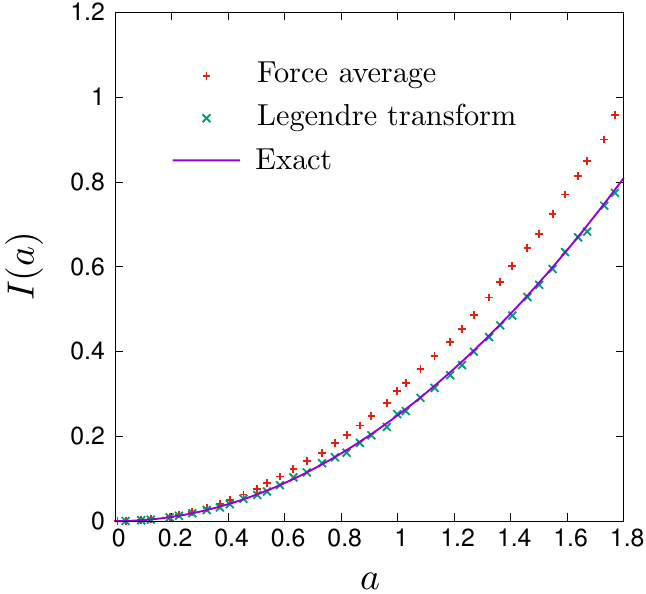}
\caption{Large deviations of the mean area for the Ornstein-Uhlenbeck process.
Left: Estimated SCGF compared with the exact result.
Right: Estimated rate function based on the Legendre transform of the SCGF and on the averaging of the biasing force (using the Hermite basis), compared with the exact result.}
\label{fig:OUlargedev}
\end{figure}

Repeating the simulations for other values of $k$, we can recover $\lambda(k)$, as
shown in the left plot of Fig.~\ref{fig:OUlargedev}. The result is in good agreement with the exact solution \eqref{eq:analytical}. Estimating $\lambda'(k)$
with \eqref{eq:ATtilde} leads to an estimated rate function, by the Legendre transform \eqref{eq:legendre},
which also agrees well with the exact rate function, shown in the right plot of Fig.~\ref{fig:OUlargedev}. 
By comparison, the rate function obtained from the time average \eqref{eq:rfct1} of the extra biasing
force (with the Hermite basis) is not as good: it lies above the exact rate function and
shows a larger offset or error as $k$ increases, which does not decrease by reducing $\Dt$ or increasing the number of basis functions. 

\begin{figure}[t]
\centering
\includegraphics[width=2.5in]{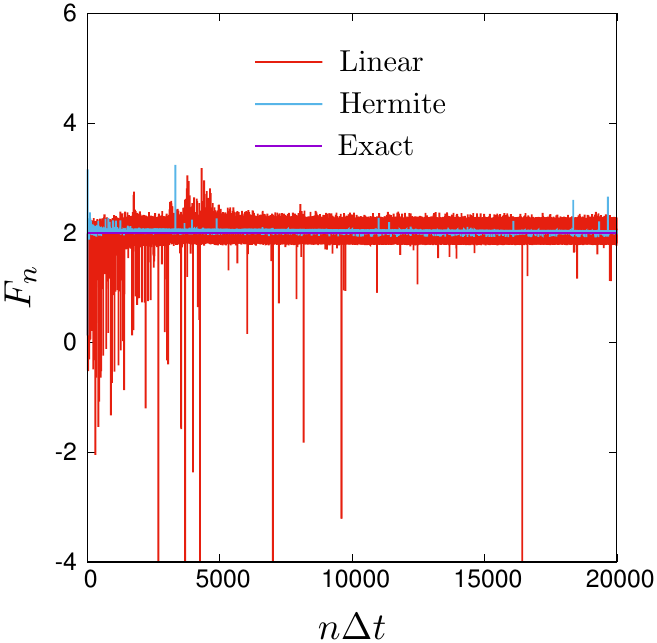}%
\hspace*{0.6in}%
\includegraphics[width=2.5in]{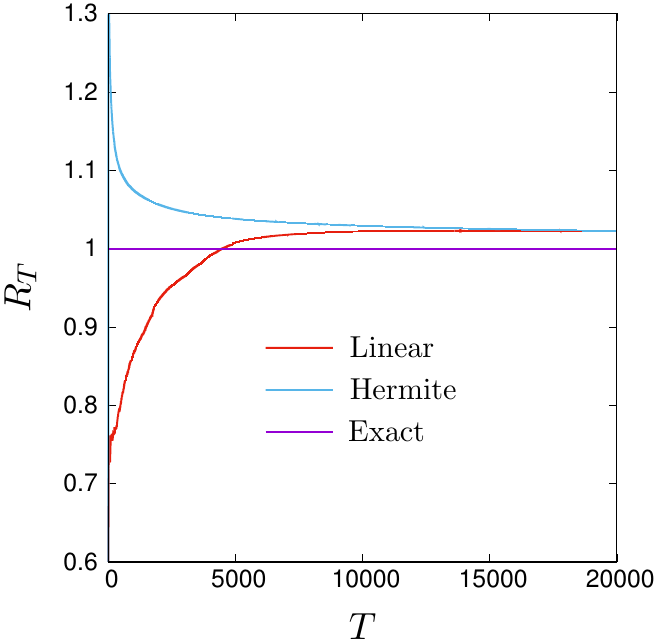}
\caption{Large deviations of the mean area of the Ornstein-Uhlenbeck process. 
Left: Biasing force $F_n$ at the current state $\x_n$ of the process at time $n$. 
Right: Evolution of the time average $R_T$, which gives the rate function according to \eqref{eq:rfct1}.}
\label{fig:OUforce}
\end{figure}

To understand this error, we show in the left plot of Fig.~\ref{fig:OUforce} the evolution of the extra biasing force $F_n$
for $k=1$, estimated at the current location of the system, which should approach the constant $2$,
following \eqref{eq:driftOU}. The evolution is noisy, as can be seen, which is expected, since $F_n$ is estimated by the logarithmic derivative
\be
\label{eq:Fratio}
F_n(x)= 2 \frac{(h^n(x))'}{h^n(x)},
\ee
computed in the function basis from \eqref{eq:Fnloc}. The derivative amplifies the Monte Carlo errors inherent in the estimate $h^n$. In
addition, the denominator often takes small values away from the mean position $m_k$, which
makes the estimation of the optimal force still more difficult. The right plot of Fig.~\ref{fig:OUforce} shows that the noise on $F_n$ is considerably filtered out by the time average underlying the estimator $R_T$ of the rate function, although a bias remains even after the convergence time of the SCGF, which leads to the offset seen in Fig.~\ref{fig:OUlargedev}. The results in both cases are more noisy for
the linear basis because the derivative is not continuous across the different cells. 

The offset on $R_T$ remains more or less constant by running independent simulations (error bars on 30 simulations are too small to show), and so appears to be a systematic error or bias. Many factors can account for this bias. First, the rate function estimated from the limit shown in \eqref{eq:rfct1} is known to be an upper bound on the true rate function \cite{chetrite2015}, which is tight if and only if the modified drift estimated in the simulation is the optimal control drift \eqref{eq:bk}. Here, $F_n$ is not constant, as predicted from \eqref{eq:driftOU}, so we expect the estimate $R_T$ to lie above its expected value. Second, we have noticed in simulations that $A_T$ underestimates $a_k$ for large $k$, which has the effect of further ``pushing'' the estimate of the rate function above $I(a)$. This is most likely due again to $F_n$ being non-constant. Finally, the time average of $F_n$ involves the ratio of two functions, according to \eqref{eq:Fratio}, which fluctuate in time via the updating of $\alpha^n$ in \eqref{eq:Fnloc}. As a result, we expect this additional randomness to artificially increase the second moment of $F_n$, leading to a further bias. 

It is difficult to isolate these factors, and all, in fact, seem to play a role. In future works, it would be interesting to study the bias observed in $R_T$ by computing, for example, the time average of $F_n^2$ using the final estimate of $h_k$ rather than the time-evolved estimate $h^n$. Different annealing sequences or averaging techniques could also be used to ``filter'' the time average of $F_n^2$ and mitigate the bias on $R_T$. For now, the most efficient and reliable way to compute the rate function is to use the estimator based on the SCGF and its Legendre transform.

\subsection{Periodic diffusion}
\label{sec:torus}

We consider for the second test a diffusion on the unit circle satisfying the SDE
\be
\label{eq:torus}
dX_t = (- V'(X_t)+\gamma)dt  + \sqrt{2}\,dW_t,
\ee
with the periodic potential $V(x)=\cos(2\pi x)$ and $\gamma\in\R$ a constant drive. For $\gamma\neq 0$, the total drift cannot be expressed as the gradient of a smooth periodic function, so $X_t$ is a nonequilibrium process violating detailed balance \cite{risken1996}. The observable studied for this process is the winding number
\be
A_T = \frac{1}{T}\int_0^T dX_t,
\ee
calculated with the real rather than periodic state, which can be interpreted physically as the mean velocity or current of a Brownian particle moving around the circle \cite{tsobgni2016}. In the notation of Sec.~\ref{sec:setting}, we have $b(x)= -V'(x) + \gamma$, $\sigma=\sqrt{2}$, $f(x)=0$, and $g(x)=1$. 

The SCGF and the rate function of this observable are not known exactly, but Galerkin approximations can easily be found by Fourier series, using the basis functions $\phi_j(x)=\e^{j 2\pi \mathrm{i} x}$ with $j\in\Z$ \cite{mehl2008,nemoto2011b,tsobgni2016}. This basis is used to compute reference values for the SCGF and the rate function, devoid of $\Delta t$ errors,
by projecting the spectral problem \eqref{eq:spectral} in Fourier space and by ensuring that enough basis
functions are used. We find in practice
that $M=41$ Fourier modes are sufficient. We use the same Fourier basis for the algorithm,
also with $M= 41$ modes, in addition to $T=2\times10^4$ and $T_1=T_2=2\times10^3$ for the integration times, as in the first test.

\begin{figure}[t]
\centering
\includegraphics[width=2.5in]{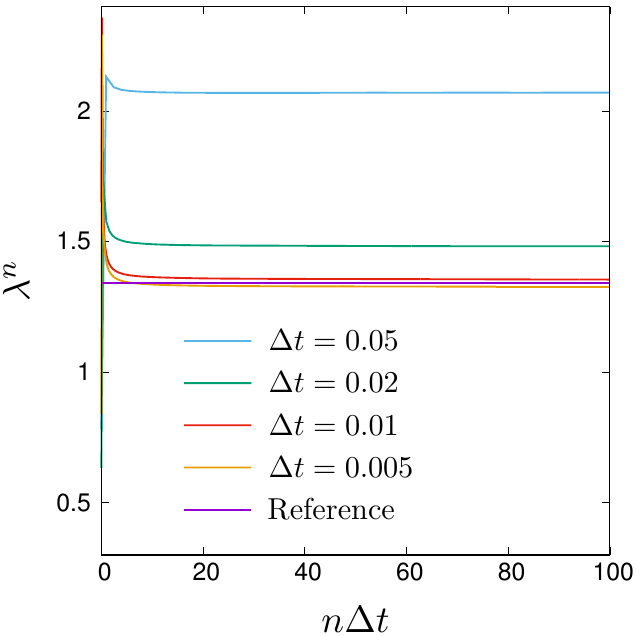}%
\hspace*{0.6in}%
\includegraphics[width=2.5in]{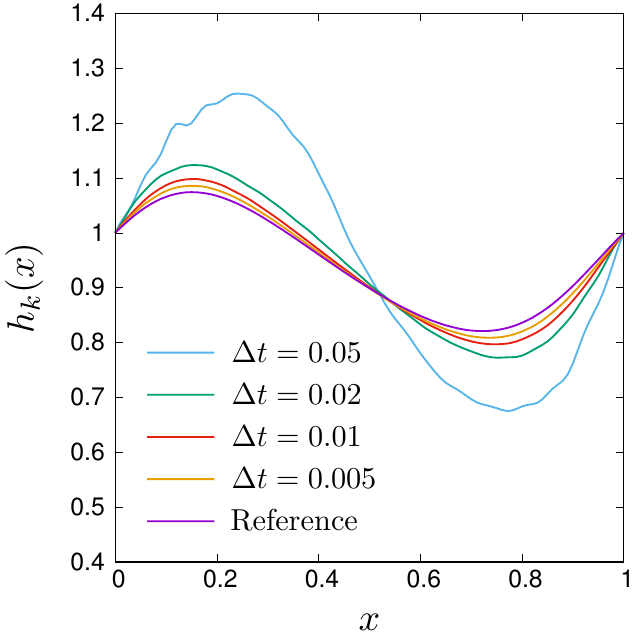}
\caption{Current large deviations of the driven periodic diffusion.
Left: Evolution of the estimated SCGF with time for different time steps $\Dt$.
Right: Estimated eigenfunction $h_k$ for different time steps $\Dt$ compared with the reference Fourier solution.}
\label{fig:toruslearning}
\end{figure}

We show in Fig.~\ref{fig:toruslearning} the results of a typical simulation for $\gamma=1$ and $k=1$. The left
plot in this figure shows the evolution in time of the estimated SCGF from the eigenvalue
iteration \eqref{eq:lambdaestimate}, while the right plot shows the final estimated
eigenfunction $h_{k}$ for different time steps $\Dt$, compared with the reference Fourier solution. We
see that the results for the eigenvalue and the eigenvector significantly depart from the reference solutions for $\Dt=0.05$, but converge
to them as $\Dt$ is decreased, in accordance with Assumption~\ref{as:spectralapprox}
and the theoretical results of \cite{ferre2017}. For the same
time step used for the Ornstein-Uhlenbeck process, namely, $\Dt=5\times 10^{-3}$, we see no
notable difference between the estimated
eigenfunction and the reference values, leading to a precise estimation of the SCGF.
The convergence of the eigenvalue here is much faster than for the Ornstein-Uhlenbeck
process because the space explored is compact, being limited to $[0,1]$ with periodic
boundary conditions.

\begin{figure}[t]
\centering
\includegraphics[width=2.5in]{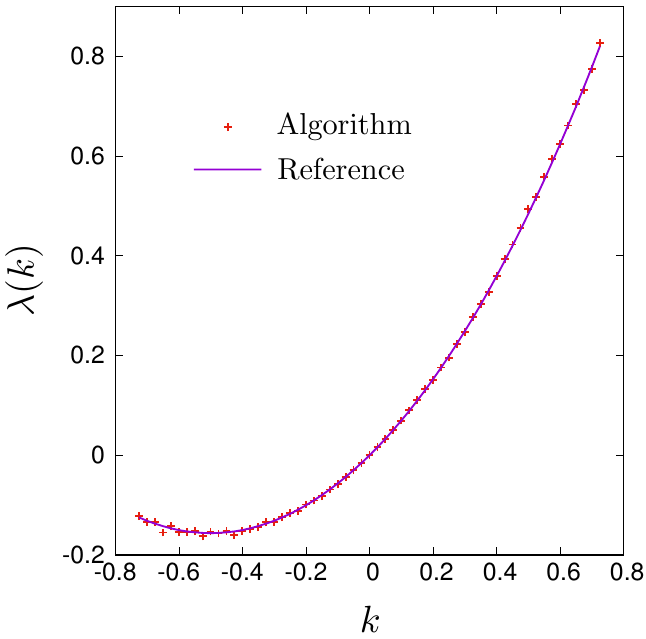}%
\hspace*{0.6in}%
\includegraphics[width=2.45in]{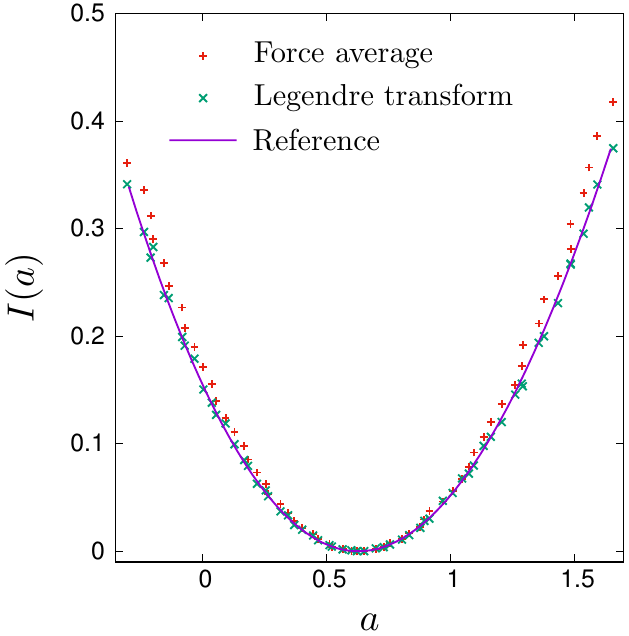}
\caption{Current large deviations for the driven periodic diffusion.
Left: Estimated SCGF compared with the reference Fourier solution.
Right: Estimated rate function based on the  Legendre transform and on the averaging of the biasing force, compared with the reference Fourier solution.}
\label{fig:toruslargedev}
\end{figure}

As for the Ornstein-Uhlenbeck process, we can repeat these simulations over a range of values for $k$ to obtain
the SCGF and the rate function. The left plot of Fig.~\ref{fig:toruslargedev} shows that the SCGF is in good agreement with the
Fourier solution for $\Dt=5\times 10^{-3}$, and so is the rate function estimated by Legendre transform. However, as seen before, the rate function estimated with the time average $R_T$ of the biasing force shows an offset, although smaller this time, which comes from the noisy estimation of $F_n$. As before, the estimator of the rate function that should be used is the one based on the Legendre transform of the SCGF, with $a_k$ estimated by $A_T$.

\section{Conclusion}

We have presented a new algorithm for estimating the large deviation functions of time-integrated observables of Markov processes, which characterize the likelihood of their fluctuations  in the long-time limit. The algorithm draws on earlier results on stochastic control \cite{borkar2004,ahamed2006,basu2008,borkar2010}, and works by adaptively estimating the principal eigenvalue and eigenfunction of a spectral problem related to the large deviation problem. The adaptive part consists in modifying the process considered, using feedback and reinforcement learning, so as to reach the so-called driven process, which is known to be optimal for the purpose of estimating large deviation functions using importance sampling \cite{chetrite2015}. In this sense, the algorithm relates to many adaptive importance sampling methods that have been proposed recently for rare event simulations; see, for instance, \cite{dupuis2005,nemoto2017b,blanchet2012,kappen2016}. It is also closely related to diffusion Monte Carlo methods \cite{foulkes2001quantum,rousset2006control,lim2017fast}, which attempt to estimate the ground-state wavefunction of a many-body quantum system by simulating a related stochastic process.

The proposed algorithm can be applied to diffusions, as illustrated here, but also to Markov chains in discrete or continuous time, and works for both equilibrium and nonequilibrium systems, that is, reversible and non-reversible processes, respectively. Moreover, although the test cases that we have presented are simple, they clearly show that the algorithm has the potential to improve upon other simulation methods, especially cloning methods, since it runs on a single simulation of the process and provides information about how fluctuations are created in time by constructing the driven process in a non-parametric form and with no prior information. A modification of the original cloning algorithm was proposed recently \cite{nemoto2016} to construct the driven process, but it is based on a different feedback rule that compares two time-dependent histograms, whose estimation is noisy and requires a large number of clones. The results presented here show that a single clone, evolving over a long-enough time, is sufficient. This obviously cuts the computational complexity of estimating large deviations, but also simplifies, as mentioned, the error and convergence analyses of the algorithm.

As with any new proposal, more work is needed to understand the benefits and limitations of the algorithm, to test its applicability to realistic systems, and to benchmark it against other
numerical methods. Of particular importance is to derive precise error estimates associated with the space and time discretizations of the spectral elements, large deviation functions, and the driven process. Recent results about the discretization errors in $\Delta t$ for the SCGF can be found in \cite{ferre2017}. These discretization errors are also present in cloning algorithms when applied to diffusions, so they are not specific to the algorithm presented here. The use of Galerkin discretizations is also common to both algorithms and requires further investigations, particularly in the low-noise limit and for processes involving many interacting particles. 

\begin{acknowledgments}
We are grateful to Florian Angeletti for useful discussions in the initial phase of this work and to Hadrien Vroylandt for kindly pointing out to us an error in Eq.~\eqref{eqsmw}, corrected in this version. G.F.\ is supported by the Labex Bezout. H.T.\ was supported by the National Research Foundation of South Africa (Grant nos 90322 and 96199) and Stellenbosch University (Project Funding for New Appointee). This research was also supported in part by the International Centre for Theoretical Sciences (ICTS) during a visit for participating in the program ``Large deviation theory in statistical physics: Recent advances and future challenges'' (Code: ICTS/Prog-ldt/2017/8).
\end{acknowledgments}

\bibliography{masterbib,complementary}

\begin{thebibliography}{67}%
\makeatletter
\providecommand \@ifxundefined [1]{%
 \@ifx{#1\undefined}
}%
\providecommand \@ifnum [1]{%
 \ifnum #1\expandafter \@firstoftwo
 \else \expandafter \@secondoftwo
 \fi
}%
\providecommand \@ifx [1]{%
 \ifx #1\expandafter \@firstoftwo
 \else \expandafter \@secondoftwo
 \fi
}%
\providecommand \natexlab [1]{#1}%
\providecommand \enquote  [1]{``#1''}%
\providecommand \bibnamefont  [1]{#1}%
\providecommand \bibfnamefont [1]{#1}%
\providecommand \citenamefont [1]{#1}%
\providecommand \href@noop [0]{\@secondoftwo}%
\providecommand \href [0]{\begingroup \@sanitize@url \@href}%
\providecommand \@href[1]{\@@startlink{#1}\@@href}%
\providecommand \@@href[1]{\endgroup#1\@@endlink}%
\providecommand \@sanitize@url [0]{\catcode `\\12\catcode `\$12\catcode
  `\&12\catcode `\#12\catcode `\^12\catcode `\_12\catcode `\%12\relax}%
\providecommand \@@startlink[1]{}%
\providecommand \@@endlink[0]{}%
\providecommand \url  [0]{\begingroup\@sanitize@url \@url }%
\providecommand \@url [1]{\endgroup\@href {#1}{\urlprefix }}%
\providecommand \urlprefix  [0]{URL }%
\providecommand \Eprint [0]{\href }%
\providecommand \doibase [0]{http://dx.doi.org/}%
\providecommand \selectlanguage [0]{\@gobble}%
\providecommand \bibinfo  [0]{\@secondoftwo}%
\providecommand \bibfield  [0]{\@secondoftwo}%
\providecommand \translation [1]{[#1]}%
\providecommand \BibitemOpen [0]{}%
\providecommand \bibitemStop [0]{}%
\providecommand \bibitemNoStop [0]{.\EOS\space}%
\providecommand \EOS [0]{\spacefactor3000\relax}%
\providecommand \BibitemShut  [1]{\csname bibitem#1\endcsname}%
\let\auto@bib@innerbib\@empty
\bibitem [{\citenamefont {Shwartz}\ and\ \citenamefont
  {Weiss}(1995)}]{shwartz1995}%
  \BibitemOpen
  \bibfield  {author} {\bibinfo {author} {\bibfnamefont {A.}~\bibnamefont
  {Shwartz}}\ and\ \bibinfo {author} {\bibfnamefont {A.}~\bibnamefont
  {Weiss}},\ }\href {http://www.crcpress.com/product/isbn/9780412063114} {\emph
  {\bibinfo {title} {Large Deviations for Performance Analysis}}},\ Stochastic
  Modeling Series\ (\bibinfo  {publisher} {Chapman and Hall},\ \bibinfo
  {address} {London},\ \bibinfo {year} {1995})\BibitemShut {NoStop}%
\bibitem [{\citenamefont {Dembo}\ and\ \citenamefont
  {Zeitouni}(1998)}]{dembo1998}%
  \BibitemOpen
  \bibfield  {author} {\bibinfo {author} {\bibfnamefont {A.}~\bibnamefont
  {Dembo}}\ and\ \bibinfo {author} {\bibfnamefont {O.}~\bibnamefont
  {Zeitouni}},\ }\href {http://www.springer.com/us/book/9783642033100} {\emph
  {\bibinfo {title} {Large Deviations Techniques and Applications}}},\ \bibinfo
  {edition} {2nd}\ ed.\ (\bibinfo  {publisher} {Springer},\ \bibinfo {address}
  {New York},\ \bibinfo {year} {1998})\BibitemShut {NoStop}%
\bibitem [{\citenamefont {{den Hollander}}(2000)}]{hollander2000}%
  \BibitemOpen
  \bibfield  {author} {\bibinfo {author} {\bibfnamefont {F.}~\bibnamefont {{den
  Hollander}}},\ }\href {http://bookstore.ams.org/fim-14.s} {\emph {\bibinfo
  {title} {Large Deviations}}},\ Fields Institute Monograph\ (\bibinfo
  {publisher} {AMS},\ \bibinfo {address} {Providence},\ \bibinfo {year}
  {2000})\BibitemShut {NoStop}%
\bibitem [{\citenamefont {Jack}\ and\ \citenamefont
  {Sollich}(2010)}]{jack2010b}%
  \BibitemOpen
  \bibfield  {author} {\bibinfo {author} {\bibfnamefont {R.~L.}\ \bibnamefont
  {Jack}}\ and\ \bibinfo {author} {\bibfnamefont {P.}~\bibnamefont {Sollich}},\
  }\bibfield  {title} {\enquote {\bibinfo {title} {Large deviations and
  ensembles of trajectories in stochastic models},}\ }\href {\doibase
  10.1143/PTPS.184.304} {\bibfield  {journal} {\bibinfo  {journal} {Prog.
  Theoret. Phys. Suppl.}\ }\textbf {\bibinfo {volume} {184}},\ \bibinfo {pages}
  {304--317} (\bibinfo {year} {2010})}\BibitemShut {NoStop}%
\bibitem [{\citenamefont {Chetrite}\ and\ \citenamefont
  {Touchette}(2013)}]{chetrite2013}%
  \BibitemOpen
  \bibfield  {author} {\bibinfo {author} {\bibfnamefont {R.}~\bibnamefont
  {Chetrite}}\ and\ \bibinfo {author} {\bibfnamefont {H.}~\bibnamefont
  {Touchette}},\ }\bibfield  {title} {\enquote {\bibinfo {title}
  {Nonequilibrium microcanonical and canonical ensembles and their
  equivalence},}\ }\href {\doibase 10.1103/PhysRevLett.111.120601} {\bibfield
  {journal} {\bibinfo  {journal} {Phys. Rev. Lett.}\ }\textbf {\bibinfo
  {volume} {111}},\ \bibinfo {pages} {120601} (\bibinfo {year}
  {2013})}\BibitemShut {NoStop}%
\bibitem [{\citenamefont {Chetrite}\ and\ \citenamefont
  {Touchette}(2015{\natexlab{a}})}]{chetrite2014}%
  \BibitemOpen
  \bibfield  {author} {\bibinfo {author} {\bibfnamefont {R.}~\bibnamefont
  {Chetrite}}\ and\ \bibinfo {author} {\bibfnamefont {H.}~\bibnamefont
  {Touchette}},\ }\bibfield  {title} {\enquote {\bibinfo {title}
  {Nonequilibrium {M}arkov processes conditioned on large deviations},}\ }\href
  {\doibase 10.1007/s00023-014-0375-8} {\bibfield  {journal} {\bibinfo
  {journal} {Ann. Henri Poincar\'e}\ }\textbf {\bibinfo {volume} {16}},\
  \bibinfo {pages} {2005--2057} (\bibinfo {year}
  {2015}{\natexlab{a}})}\BibitemShut {NoStop}%
\bibitem [{\citenamefont {Touchette}(2009)}]{touchette2009}%
  \BibitemOpen
  \bibfield  {author} {\bibinfo {author} {\bibfnamefont {H.}~\bibnamefont
  {Touchette}},\ }\bibfield  {title} {\enquote {\bibinfo {title} {The large
  deviation approach to statistical mechanics},}\ }\href {\doibase
  10.1016/j.physrep.2009.05.002} {\bibfield  {journal} {\bibinfo  {journal}
  {Phys. Rep.}\ }\textbf {\bibinfo {volume} {478}},\ \bibinfo {pages} {1--69}
  (\bibinfo {year} {2009})}\BibitemShut {NoStop}%
\bibitem [{\citenamefont {Derrida}(2007)}]{derrida2007}%
  \BibitemOpen
  \bibfield  {author} {\bibinfo {author} {\bibfnamefont {B.}~\bibnamefont
  {Derrida}},\ }\bibfield  {title} {\enquote {\bibinfo {title} {Non-equilibrium
  steady states: {F}luctuations and large deviations of the density and of the
  current},}\ }\href {\doibase 10.1088/1742-5468/2007/07/P07023} {\bibfield
  {journal} {\bibinfo  {journal} {J. Stat. Mech.}\ }\textbf {\bibinfo {volume}
  {2007}},\ \bibinfo {pages} {P07023} (\bibinfo {year} {2007})}\BibitemShut
  {NoStop}%
\bibitem [{\citenamefont {Harris}\ and\ \citenamefont
  {Touchette}(2013)}]{harris2013}%
  \BibitemOpen
  \bibfield  {author} {\bibinfo {author} {\bibfnamefont {R.~J.}\ \bibnamefont
  {Harris}}\ and\ \bibinfo {author} {\bibfnamefont {H.}~\bibnamefont
  {Touchette}},\ }\bibfield  {title} {\enquote {\bibinfo {title} {Large
  deviation approach to nonequilibrium systems},}\ }in\ \href {\doibase
  10.1002/9783527658701.ch11} {\emph {\bibinfo {booktitle} {Nonequilibrium
  Statistical Physics of Small Systems: {F}luctuation Relations and Beyond}}},\
  \bibinfo {series} {Reviews of Nonlinear Dynamics and Complexity},
  Vol.~\bibinfo {volume} {6},\ \bibinfo {editor} {edited by\ \bibinfo {editor}
  {\bibfnamefont {R.}~\bibnamefont {Klages}}, \bibinfo {editor} {\bibfnamefont
  {W.}~\bibnamefont {Just}}, \ and\ \bibinfo {editor} {\bibfnamefont
  {C.}~\bibnamefont {Jarzynski}}}\ (\bibinfo  {publisher} {Wiley-VCH},\
  \bibinfo {address} {Weinheim},\ \bibinfo {year} {2013})\ pp.\ \bibinfo
  {pages} {335--360}\BibitemShut {NoStop}%
\bibitem [{\citenamefont {Harris}\ and\ \citenamefont
  {Sch\"{u}tz}(2007)}]{harris2007}%
  \BibitemOpen
  \bibfield  {author} {\bibinfo {author} {\bibfnamefont {R.~J.}\ \bibnamefont
  {Harris}}\ and\ \bibinfo {author} {\bibfnamefont {G.~M.}\ \bibnamefont
  {Sch\"{u}tz}},\ }\bibfield  {title} {\enquote {\bibinfo {title} {Fluctuation
  theorems for stochastic dynamics},}\ }\href {\doibase
  10.1088/1742-5468/2007/07/P07020} {\bibfield  {journal} {\bibinfo  {journal}
  {J. Stat. Mech.}\ }\textbf {\bibinfo {volume} {2007}},\ \bibinfo {pages}
  {P07020} (\bibinfo {year} {2007})}\BibitemShut {NoStop}%
\bibitem [{\citenamefont {Garrahan}\ \emph {et~al.}(2007)\citenamefont
  {Garrahan}, \citenamefont {Jack}, \citenamefont {Lecomte}, \citenamefont
  {Pitard}, \citenamefont {van Duijvendijk},\ and\ \citenamefont {van
  Wijland}}]{garrahan2007}%
  \BibitemOpen
  \bibfield  {author} {\bibinfo {author} {\bibfnamefont {J.~P.}\ \bibnamefont
  {Garrahan}}, \bibinfo {author} {\bibfnamefont {R.~L.}\ \bibnamefont {Jack}},
  \bibinfo {author} {\bibfnamefont {V.}~\bibnamefont {Lecomte}}, \bibinfo
  {author} {\bibfnamefont {E.}~\bibnamefont {Pitard}}, \bibinfo {author}
  {\bibfnamefont {K.}~\bibnamefont {van Duijvendijk}}, \ and\ \bibinfo {author}
  {\bibfnamefont {F.}~\bibnamefont {van Wijland}},\ }\bibfield  {title}
  {\enquote {\bibinfo {title} {Dynamical first-order phase transition in
  kinetically constrained models of glasses},}\ }\href {\doibase
  10.1103/PhysRevLett.98.195702} {\bibfield  {journal} {\bibinfo  {journal}
  {Phys. Rev. Lett.}\ }\textbf {\bibinfo {volume} {98}},\ \bibinfo {pages}
  {195702} (\bibinfo {year} {2007})}\BibitemShut {NoStop}%
\bibitem [{\citenamefont {Hedges}\ \emph {et~al.}(2009)\citenamefont {Hedges},
  \citenamefont {Jack}, \citenamefont {Garrahan},\ and\ \citenamefont
  {Chandler}}]{hedges2009}%
  \BibitemOpen
  \bibfield  {author} {\bibinfo {author} {\bibfnamefont {L.~O.}\ \bibnamefont
  {Hedges}}, \bibinfo {author} {\bibfnamefont {R.~L.}\ \bibnamefont {Jack}},
  \bibinfo {author} {\bibfnamefont {J.~P.}\ \bibnamefont {Garrahan}}, \ and\
  \bibinfo {author} {\bibfnamefont {D.}~\bibnamefont {Chandler}},\ }\bibfield
  {title} {\enquote {\bibinfo {title} {Dynamic order-disorder in atomistic
  models of structural glass formers},}\ }\href {\doibase
  10.1126/science.1166665} {\bibfield  {journal} {\bibinfo  {journal}
  {Science}\ }\textbf {\bibinfo {volume} {323}},\ \bibinfo {pages} {1309--1313}
  (\bibinfo {year} {2009})}\BibitemShut {NoStop}%
\bibitem [{\citenamefont {Espigares}\ \emph {et~al.}(2013)\citenamefont
  {Espigares}, \citenamefont {Garrido},\ and\ \citenamefont
  {Hurtado}}]{espigares2013}%
  \BibitemOpen
  \bibfield  {author} {\bibinfo {author} {\bibfnamefont {C.~P.}\ \bibnamefont
  {Espigares}}, \bibinfo {author} {\bibfnamefont {P.~L.}\ \bibnamefont
  {Garrido}}, \ and\ \bibinfo {author} {\bibfnamefont {P.~I.}\ \bibnamefont
  {Hurtado}},\ }\bibfield  {title} {\enquote {\bibinfo {title} {Dynamical phase
  transition for current statistics in a simple driven diffusive system},}\
  }\href {\doibase 10.1103/PhysRevE.87.032115} {\bibfield  {journal} {\bibinfo
  {journal} {Phys. Rev. E}\ }\textbf {\bibinfo {volume} {87}},\ \bibinfo
  {pages} {032115} (\bibinfo {year} {2013})}\BibitemShut {NoStop}%
\bibitem [{\citenamefont {Aminov}\ \emph {et~al.}(2014)\citenamefont {Aminov},
  \citenamefont {Bunin},\ and\ \citenamefont {Kafri}}]{aminov2014}%
  \BibitemOpen
  \bibfield  {author} {\bibinfo {author} {\bibfnamefont {A.}~\bibnamefont
  {Aminov}}, \bibinfo {author} {\bibfnamefont {G.}~\bibnamefont {Bunin}}, \
  and\ \bibinfo {author} {\bibfnamefont {Y.}~\bibnamefont {Kafri}},\ }\bibfield
   {title} {\enquote {\bibinfo {title} {Singularities in large deviation
  functionals of bulk-driven transport models},}\ }\href
  {http://stacks.iop.org/1742-5468/2014/i=8/a=P08017} {\bibfield  {journal}
  {\bibinfo  {journal} {J. Stat. Mech.}\ }\textbf {\bibinfo {volume} {2014}},\
  \bibinfo {pages} {P08017} (\bibinfo {year} {2014})}\BibitemShut {NoStop}%
\bibitem [{\citenamefont {Dean}\ and\ \citenamefont {Dupuis}(2009)}]{dean2009}%
  \BibitemOpen
  \bibfield  {author} {\bibinfo {author} {\bibfnamefont {T.}~\bibnamefont
  {Dean}}\ and\ \bibinfo {author} {\bibfnamefont {P.}~\bibnamefont {Dupuis}},\
  }\bibfield  {title} {\enquote {\bibinfo {title} {Splitting for rare event
  simulation: {A} large deviation approach to design and analysis},}\ }\href
  {\doibase 10.1016/j.spa.2008.02.017} {\bibfield  {journal} {\bibinfo
  {journal} {Stoch. Proc. Appl.}\ }\textbf {\bibinfo {volume} {119}},\ \bibinfo
  {pages} {562--587} (\bibinfo {year} {2009})}\BibitemShut {NoStop}%
\bibitem [{\citenamefont {C\'erou}\ and\ \citenamefont
  {Guyader}(2007)}]{cerou2007}%
  \BibitemOpen
  \bibfield  {author} {\bibinfo {author} {\bibfnamefont {F.}~\bibnamefont
  {C\'erou}}\ and\ \bibinfo {author} {\bibfnamefont {A.}~\bibnamefont
  {Guyader}},\ }\bibfield  {title} {\enquote {\bibinfo {title} {Adaptive
  multilevel splitting for rare event analysis},}\ }\href {\doibase
  10.1080/07362990601139628} {\bibfield  {journal} {\bibinfo  {journal} {Stoch.
  Anal. Appl.}\ }\textbf {\bibinfo {volume} {25}},\ \bibinfo {pages} {417--443}
  (\bibinfo {year} {2007})}\BibitemShut {NoStop}%
\bibitem [{\citenamefont {Aristoff}\ \emph {et~al.}(2015)\citenamefont
  {Aristoff}, \citenamefont {Leli\`evre}, \citenamefont {Mayne},\ and\
  \citenamefont {Teo}}]{aristoff2015}%
  \BibitemOpen
  \bibfield  {author} {\bibinfo {author} {\bibfnamefont {D.}~\bibnamefont
  {Aristoff}}, \bibinfo {author} {\bibfnamefont {T.}~\bibnamefont
  {Leli\`evre}}, \bibinfo {author} {\bibfnamefont {C.~G.}\ \bibnamefont
  {Mayne}}, \ and\ \bibinfo {author} {\bibfnamefont {I.}~\bibnamefont {Teo}},\
  }\bibfield  {title} {\enquote {\bibinfo {title} {Adaptive multilevel
  splitting in molecular dynamics simulations},}\ }\href {\doibase
  10.1051/proc/201448009} {\bibfield  {journal} {\bibinfo  {journal} {ESAIM:
  Proc.}\ }\textbf {\bibinfo {volume} {48}},\ \bibinfo {pages} {215--225}
  (\bibinfo {year} {2015})}\BibitemShut {NoStop}%
\bibitem [{\citenamefont {Grassberger}(2002)}]{grassberger2002}%
  \BibitemOpen
  \bibfield  {author} {\bibinfo {author} {\bibfnamefont {P.}~\bibnamefont
  {Grassberger}},\ }\bibfield  {title} {\enquote {\bibinfo {title} {Go with the
  winners: {A} general {M}onte {C}arlo strategy},}\ }\href {\doibase
  10.1016/S0010-4655(02)00205-9} {\bibfield  {journal} {\bibinfo  {journal}
  {Comp. Phys. Comm.}\ }\textbf {\bibinfo {volume} {147}},\ \bibinfo {pages}
  {64--70} (\bibinfo {year} {2002})}\BibitemShut {NoStop}%
\bibitem [{\citenamefont {Giardina}\ \emph {et~al.}(2006)\citenamefont
  {Giardina}, \citenamefont {Kurchan},\ and\ \citenamefont
  {Peliti}}]{giardina2006}%
  \BibitemOpen
  \bibfield  {author} {\bibinfo {author} {\bibfnamefont {C.}~\bibnamefont
  {Giardina}}, \bibinfo {author} {\bibfnamefont {J.}~\bibnamefont {Kurchan}}, \
  and\ \bibinfo {author} {\bibfnamefont {L.}~\bibnamefont {Peliti}},\
  }\bibfield  {title} {\enquote {\bibinfo {title} {Direct evaluation of
  large-deviation functions},}\ }\href
  {http://link.aps.org/abstract/PRL/v96/e120603} {\bibfield  {journal}
  {\bibinfo  {journal} {Phys. Rev. Lett.}\ }\textbf {\bibinfo {volume} {96}},\
  \bibinfo {eid} {120603} (\bibinfo {year} {2006})}\BibitemShut {NoStop}%
\bibitem [{\citenamefont {Lecomte}\ and\ \citenamefont
  {Tailleur}(2007)}]{lecomte2007a}%
  \BibitemOpen
  \bibfield  {author} {\bibinfo {author} {\bibfnamefont {V.}~\bibnamefont
  {Lecomte}}\ and\ \bibinfo {author} {\bibfnamefont {J.}~\bibnamefont
  {Tailleur}},\ }\bibfield  {title} {\enquote {\bibinfo {title} {A numerical
  approach to large deviations in continuous time},}\ }\href {\doibase
  10.1088/1742-5468/2007/03/P03004} {\bibfield  {journal} {\bibinfo  {journal}
  {J. Stat. Mech.}\ }\textbf {\bibinfo {volume} {2007}},\ \bibinfo {pages}
  {P03004} (\bibinfo {year} {2007})}\BibitemShut {NoStop}%
\bibitem [{\citenamefont {Bucklew}(2004)}]{bucklew2004}%
  \BibitemOpen
  \bibfield  {author} {\bibinfo {author} {\bibfnamefont {J.~A.}\ \bibnamefont
  {Bucklew}},\ }\href
  {http://www.springer.com/statistics/physical+%26+information+science/book/978-0-387-20078-1}
  {\emph {\bibinfo {title} {Introduction to Rare Event Simulation}}}\ (\bibinfo
   {publisher} {Springer},\ \bibinfo {address} {New York},\ \bibinfo {year}
  {2004})\BibitemShut {NoStop}%
\bibitem [{\citenamefont {Juneja}\ and\ \citenamefont
  {Shahabuddin}(2006)}]{juneja2006}%
  \BibitemOpen
  \bibfield  {author} {\bibinfo {author} {\bibfnamefont {S.}~\bibnamefont
  {Juneja}}\ and\ \bibinfo {author} {\bibfnamefont {P.}~\bibnamefont
  {Shahabuddin}},\ }\enquote {\bibinfo {title} {Rare-event simulation
  techniques: An introduction and recent advances},}\ in\ \href {\doibase
  http://dx.doi.org/10.1016/S0927-0507(06)13011-X} {\emph {\bibinfo {booktitle}
  {Handbooks in Operations Research and Management Science}}},\ Vol.\ \bibinfo
  {volume} {Volume 13},\ \bibinfo {editor} {edited by\ \bibinfo {editor}
  {\bibfnamefont {S.~G.}\ \bibnamefont {Henderson}}\ and\ \bibinfo {editor}
  {\bibfnamefont {B.~L.}\ \bibnamefont {Nelson}}}\ (\bibinfo  {publisher}
  {Elsevier},\ \bibinfo {address} {Amsterdam},\ \bibinfo {year} {2006})\
  Chap.~\bibinfo {chapter} {11}, pp.\ \bibinfo {pages} {291--350}\BibitemShut
  {NoStop}%
\bibitem [{\citenamefont {Asmussen}\ and\ \citenamefont
  {Glynn}(2007)}]{asmussen2007}%
  \BibitemOpen
  \bibfield  {author} {\bibinfo {author} {\bibfnamefont {S.}~\bibnamefont
  {Asmussen}}\ and\ \bibinfo {author} {\bibfnamefont {P.~W.}\ \bibnamefont
  {Glynn}},\ }\href
  {http://www.springer.com/mathematics/probability/book/978-0-387-30679-7}
  {\emph {\bibinfo {title} {Stochastic Simulation: Algorithms and Analysis}}},\
  Stochastic Modelling and Applied Probability\ (\bibinfo  {publisher}
  {Springer},\ \bibinfo {address} {New York},\ \bibinfo {year}
  {2007})\BibitemShut {NoStop}%
\bibitem [{\citenamefont {Bolhuis}\ \emph {et~al.}(2002)\citenamefont
  {Bolhuis}, \citenamefont {Chandler}, \citenamefont {Dellago},\ and\
  \citenamefont {Geissler}}]{bolhuis2002}%
  \BibitemOpen
  \bibfield  {author} {\bibinfo {author} {\bibfnamefont {P.~G.}\ \bibnamefont
  {Bolhuis}}, \bibinfo {author} {\bibfnamefont {D.}~\bibnamefont {Chandler}},
  \bibinfo {author} {\bibfnamefont {C.}~\bibnamefont {Dellago}}, \ and\
  \bibinfo {author} {\bibfnamefont {P.~L.}\ \bibnamefont {Geissler}},\
  }\bibfield  {title} {\enquote {\bibinfo {title} {Transition path sampling:
  {T}hrowing ropes over rough mountain passes, in the dark},}\ }\href {\doibase
  10.1146/annurev.physchem.53.082301.113146} {\bibfield  {journal} {\bibinfo
  {journal} {Ann. Rev. Phys. Chem.}\ }\textbf {\bibinfo {volume} {53}},\
  \bibinfo {pages} {291--318} (\bibinfo {year} {2002})}\BibitemShut {NoStop}%
\bibitem [{\citenamefont {Heymann}\ and\ \citenamefont {{V}anden
  {E}ijnden}(2008)}]{heymann2008}%
  \BibitemOpen
  \bibfield  {author} {\bibinfo {author} {\bibfnamefont {M.}~\bibnamefont
  {Heymann}}\ and\ \bibinfo {author} {\bibfnamefont {E.}~\bibnamefont {{V}anden
  {E}ijnden}},\ }\bibfield  {title} {\enquote {\bibinfo {title} {Pathways of
  maximum likelihood for rare events in nonequilibrium systems: Application to
  nucleation in the presence of shear},}\ }\href {\doibase
  10.1103/PhysRevLett.100.140601} {\bibfield  {journal} {\bibinfo  {journal}
  {Phys. Rev. Lett.}\ }\textbf {\bibinfo {volume} {100}},\ \bibinfo {pages}
  {140601} (\bibinfo {year} {2008})}\BibitemShut {NoStop}%
\bibitem [{\citenamefont {Vanden-Eijnden}\ and\ \citenamefont
  {Weare}(2012)}]{eijnden2012}%
  \BibitemOpen
  \bibfield  {author} {\bibinfo {author} {\bibfnamefont {E.}~\bibnamefont
  {Vanden-Eijnden}}\ and\ \bibinfo {author} {\bibfnamefont {J.}~\bibnamefont
  {Weare}},\ }\bibfield  {title} {\enquote {\bibinfo {title} {Rare event
  simulation of small noise diffusions},}\ }\href {\doibase 10.1002/cpa.21428}
  {\bibfield  {journal} {\bibinfo  {journal} {Comm. Pure Appl. Math.}\ }\textbf
  {\bibinfo {volume} {65}},\ \bibinfo {pages} {1770--1803} (\bibinfo {year}
  {2012})}\BibitemShut {NoStop}%
\bibitem [{\citenamefont {Grafke}\ \emph {et~al.}(2015)\citenamefont {Grafke},
  \citenamefont {Grauer},\ and\ \citenamefont {Sch\"afer}}]{grafke2015}%
  \BibitemOpen
  \bibfield  {author} {\bibinfo {author} {\bibfnamefont {T.}~\bibnamefont
  {Grafke}}, \bibinfo {author} {\bibfnamefont {R.}~\bibnamefont {Grauer}}, \
  and\ \bibinfo {author} {\bibfnamefont {T.}~\bibnamefont {Sch\"afer}},\
  }\bibfield  {title} {\enquote {\bibinfo {title} {The instanton method and its
  numerical implementation in fluid mechanics},}\ }\href
  {http://stacks.iop.org/1751-8121/48/i=33/a=333001} {\bibfield  {journal}
  {\bibinfo  {journal} {J. Phys. A: Math. Theor.}\ }\textbf {\bibinfo {volume}
  {48}},\ \bibinfo {pages} {333001} (\bibinfo {year} {2015})}\BibitemShut
  {NoStop}%
\bibitem [{\citenamefont {Borkar}\ \emph {et~al.}(2004)\citenamefont {Borkar},
  \citenamefont {Juneja},\ and\ \citenamefont {Kherani}}]{borkar2004}%
  \BibitemOpen
  \bibfield  {author} {\bibinfo {author} {\bibfnamefont {V.~S.}\ \bibnamefont
  {Borkar}}, \bibinfo {author} {\bibfnamefont {S.}~\bibnamefont {Juneja}}, \
  and\ \bibinfo {author} {\bibfnamefont {A.~A.}\ \bibnamefont {Kherani}},\
  }\bibfield  {title} {\enquote {\bibinfo {title} {Peformance analysis
  conditioned on rare events: {A}n adaptive simulation scheme},}\ }\href
  {http://projecteuclid.org/euclid.cis/1119639799} {\bibfield  {journal}
  {\bibinfo  {journal} {Commun. Info. Syst.}\ }\textbf {\bibinfo {volume}
  {3}},\ \bibinfo {pages} {259--278} (\bibinfo {year} {2004})}\BibitemShut
  {NoStop}%
\bibitem [{\citenamefont {Ahamed}\ \emph {et~al.}(2006)\citenamefont {Ahamed},
  \citenamefont {Borkar},\ and\ \citenamefont {Juneja}}]{ahamed2006}%
  \BibitemOpen
  \bibfield  {author} {\bibinfo {author} {\bibfnamefont {T.~P.~I.}\
  \bibnamefont {Ahamed}}, \bibinfo {author} {\bibfnamefont {V.~S.}\
  \bibnamefont {Borkar}}, \ and\ \bibinfo {author} {\bibfnamefont
  {S.}~\bibnamefont {Juneja}},\ }\bibfield  {title} {\enquote {\bibinfo {title}
  {Adaptive importance sampling technique for {M}arkov chains using stochastic
  approximation},}\ }\href {\doibase 10.1287/opre.1060.0291} {\bibfield
  {journal} {\bibinfo  {journal} {Oper. Res.}\ }\textbf {\bibinfo {volume}
  {54}},\ \bibinfo {pages} {489--504} (\bibinfo {year} {2006})}\BibitemShut
  {NoStop}%
\bibitem [{\citenamefont {Basu}\ \emph {et~al.}(2008)\citenamefont {Basu},
  \citenamefont {Bhattacharyya},\ and\ \citenamefont {Borkar}}]{basu2008}%
  \BibitemOpen
  \bibfield  {author} {\bibinfo {author} {\bibfnamefont {A.}~\bibnamefont
  {Basu}}, \bibinfo {author} {\bibfnamefont {T.}~\bibnamefont {Bhattacharyya}},
  \ and\ \bibinfo {author} {\bibfnamefont {V.~S.}\ \bibnamefont {Borkar}},\
  }\bibfield  {title} {\enquote {\bibinfo {title} {A learning algorithm for
  risk-sensitive cost},}\ }\href {\doibase 10.1287/moor.1080.0324} {\bibfield
  {journal} {\bibinfo  {journal} {Math. Oper. Res.}\ }\textbf {\bibinfo
  {volume} {33}},\ \bibinfo {pages} {880--898} (\bibinfo {year}
  {2008})}\BibitemShut {NoStop}%
\bibitem [{\citenamefont {Borkar}(2010)}]{borkar2010}%
  \BibitemOpen
  \bibfield  {author} {\bibinfo {author} {\bibfnamefont {V.~S.}\ \bibnamefont
  {Borkar}},\ }\bibfield  {title} {\enquote {\bibinfo {title} {Learning
  algorithms for risk-sensitive control},}\ }in\ \href@noop {} {\emph {\bibinfo
  {booktitle} {Proc. 19th Int. Symp. Math. Theory Networks and Systems}}}\
  (\bibinfo {year} {2010})\ pp.\ \bibinfo {pages} {1327--1332}\BibitemShut
  {NoStop}%
\bibitem [{\citenamefont {Chetrite}\ and\ \citenamefont
  {Touchette}(2015{\natexlab{b}})}]{chetrite2015}%
  \BibitemOpen
  \bibfield  {author} {\bibinfo {author} {\bibfnamefont {R.}~\bibnamefont
  {Chetrite}}\ and\ \bibinfo {author} {\bibfnamefont {H.}~\bibnamefont
  {Touchette}},\ }\bibfield  {title} {\enquote {\bibinfo {title} {Variational
  and optimal control representations of conditioned and driven processes},}\
  }\href {\doibase 10.1088/1742-5468/2015/12/P12001} {\bibfield  {journal}
  {\bibinfo  {journal} {J. Stat. Mech.}\ }\textbf {\bibinfo {volume} {2015}},\
  \bibinfo {pages} {P12001} (\bibinfo {year} {2015}{\natexlab{b}})}\BibitemShut
  {NoStop}%
\bibitem [{\citenamefont {Chauveau}\ and\ \citenamefont
  {Diebolt}(2003)}]{chauveau2003estimation}%
  \BibitemOpen
  \bibfield  {author} {\bibinfo {author} {\bibfnamefont {D.}~\bibnamefont
  {Chauveau}}\ and\ \bibinfo {author} {\bibfnamefont {J.}~\bibnamefont
  {Diebolt}},\ }\bibfield  {title} {\enquote {\bibinfo {title} {Estimation of
  the asymptotic variance in the {CLT} for {M}arkov chains},}\ }\href {\doibase
  10.1081/STM-120025399} {\bibfield  {journal} {\bibinfo  {journal} {Stoch.
  Models}\ }\textbf {\bibinfo {volume} {19}},\ \bibinfo {pages} {449--465}
  (\bibinfo {year} {2003})}\BibitemShut {NoStop}%
\bibitem [{\citenamefont {Roberts}\ and\ \citenamefont
  {Rosenthal}(2004)}]{roberts2004general}%
  \BibitemOpen
  \bibfield  {author} {\bibinfo {author} {\bibfnamefont {G.~O.}\ \bibnamefont
  {Roberts}}\ and\ \bibinfo {author} {\bibfnamefont {J.~S.}\ \bibnamefont
  {Rosenthal}},\ }\bibfield  {title} {\enquote {\bibinfo {title} {General state
  space {M}arkov chains and {MCMC} algorithms},}\ }\href {\doibase
  10.1214/154957804100000024} {\bibfield  {journal} {\bibinfo  {journal} {Prob.
  Surveys}\ }\textbf {\bibinfo {volume} {1}},\ \bibinfo {pages} {20--71}
  (\bibinfo {year} {2004})}\BibitemShut {NoStop}%
\bibitem [{\citenamefont {Benveniste}\ \emph {et~al.}(2012)\citenamefont
  {Benveniste}, \citenamefont {M{\'e}tivier},\ and\ \citenamefont
  {Priouret}}]{benveniste2012adaptive}%
  \BibitemOpen
  \bibfield  {author} {\bibinfo {author} {\bibfnamefont {A.}~\bibnamefont
  {Benveniste}}, \bibinfo {author} {\bibfnamefont {M.}~\bibnamefont
  {M{\'e}tivier}}, \ and\ \bibinfo {author} {\bibfnamefont {P.}~\bibnamefont
  {Priouret}},\ }\href {http://www.springer.com/gp/book/9783642758966} {\emph
  {\bibinfo {title} {Adaptive Algorithms and Stochastic Approximations}}},\
  \bibinfo {series} {Stochastic Modelling and Applied Probability},
  Vol.~\bibinfo {volume} {22}\ (\bibinfo  {publisher} {Springer},\ \bibinfo
  {year} {2012})\BibitemShut {NoStop}%
\bibitem [{\citenamefont {Ferr\'e}\ and\ \citenamefont
  {Stoltz}(2017)}]{ferre2017}%
  \BibitemOpen
  \bibfield  {author} {\bibinfo {author} {\bibfnamefont {G.}~\bibnamefont
  {Ferr\'e}}\ and\ \bibinfo {author} {\bibfnamefont {G.}~\bibnamefont
  {Stoltz}},\ }\bibfield  {title} {\enquote {\bibinfo {title} {Error estimates
  on ergodic properties of discretized {F}eynman--{K}ac semigroups},}\
  }\href@noop {} {\  (\bibinfo {year} {2017})},\ \Eprint
  {http://arxiv.org/abs/arXiv:1712.04013} {arXiv:1712.04013} \BibitemShut
  {NoStop}%
\bibitem [{\citenamefont {Pavliotis}(2014)}]{pavliotis2014}%
  \BibitemOpen
  \bibfield  {author} {\bibinfo {author} {\bibfnamefont {G.~A.}\ \bibnamefont
  {Pavliotis}},\ }\href
  {http://link.springer.com/book/10.1007%2F978-1-4939-1323-7} {\emph {\bibinfo
  {title} {Stochastic Processes and Applications}}}\ (\bibinfo  {publisher}
  {Springer},\ \bibinfo {address} {New York},\ \bibinfo {year}
  {2014})\BibitemShut {NoStop}%
\bibitem [{\citenamefont {Chernyak}\ \emph {et~al.}(2014)\citenamefont
  {Chernyak}, \citenamefont {Chertkov}, \citenamefont {Bierkens},\ and\
  \citenamefont {Kappen}}]{chernyak2014}%
  \BibitemOpen
  \bibfield  {author} {\bibinfo {author} {\bibfnamefont {V.~Y.}\ \bibnamefont
  {Chernyak}}, \bibinfo {author} {\bibfnamefont {M.}~\bibnamefont {Chertkov}},
  \bibinfo {author} {\bibfnamefont {J.}~\bibnamefont {Bierkens}}, \ and\
  \bibinfo {author} {\bibfnamefont {H.~J.}\ \bibnamefont {Kappen}},\ }\bibfield
   {title} {\enquote {\bibinfo {title} {Stochastic optimal control as
  non-equilibrium statistical mechanics: {C}alculus of variations over density
  and current},}\ }\href {\doibase 10.1088/1751-8113/47/2/022001} {\bibfield
  {journal} {\bibinfo  {journal} {J. Phys. A: Math. Theor.}\ }\textbf {\bibinfo
  {volume} {47}},\ \bibinfo {pages} {022001} (\bibinfo {year}
  {2014})}\BibitemShut {NoStop}%
\bibitem [{\citenamefont {Sekimoto}(2010)}]{sekimoto2010}%
  \BibitemOpen
  \bibfield  {author} {\bibinfo {author} {\bibfnamefont {K.}~\bibnamefont
  {Sekimoto}},\ }\href {http://www.springer.com/us/book/9783642054105} {\emph
  {\bibinfo {title} {Stochastic Energetics}}},\ \bibinfo {series} {Lect. Notes.
  Phys.}, Vol.\ \bibinfo {volume} {799}\ (\bibinfo  {publisher} {Springer},\
  \bibinfo {address} {New York},\ \bibinfo {year} {2010})\BibitemShut {NoStop}%
\bibitem [{\citenamefont {Bierkens}\ \emph {et~al.}(2016)\citenamefont
  {Bierkens}, \citenamefont {Chernyak}, \citenamefont {Chertkov},\ and\
  \citenamefont {Kappen}}]{bierkens2013}%
  \BibitemOpen
  \bibfield  {author} {\bibinfo {author} {\bibfnamefont {J.}~\bibnamefont
  {Bierkens}}, \bibinfo {author} {\bibfnamefont {V.~Y.}\ \bibnamefont
  {Chernyak}}, \bibinfo {author} {\bibfnamefont {M.}~\bibnamefont {Chertkov}},
  \ and\ \bibinfo {author} {\bibfnamefont {H.~J.}\ \bibnamefont {Kappen}},\
  }\bibfield  {title} {\enquote {\bibinfo {title} {Linear {PDE}s and eigenvalue
  problems corresponding to ergodic stochastic optimization problems on compact
  manifolds},}\ }\href {\doibase 10.1088/1742-5468/2016/01/013206} {\bibfield
  {journal} {\bibinfo  {journal} {J. Stat. Mech.}\ }\textbf {\bibinfo {volume}
  {2016}},\ \bibinfo {pages} {013206} (\bibinfo {year} {2016})}\BibitemShut
  {NoStop}%
\bibitem [{\citenamefont {Leli\`evre}\ and\ \citenamefont
  {Stoltz}(2016)}]{lelievre2016}%
  \BibitemOpen
  \bibfield  {author} {\bibinfo {author} {\bibfnamefont {T.}~\bibnamefont
  {Leli\`evre}}\ and\ \bibinfo {author} {\bibfnamefont {G.}~\bibnamefont
  {Stoltz}},\ }\bibfield  {title} {\enquote {\bibinfo {title} {Partial
  differential equations and stochastic methods in molecular dynamics},}\
  }\href {\doibase 10.1017/S0962492916000039} {\bibfield  {journal} {\bibinfo
  {journal} {Acta Numer.}\ }\textbf {\bibinfo {volume} {25}},\ \bibinfo {pages}
  {681--880} (\bibinfo {year} {2016})}\BibitemShut {NoStop}%
\bibitem [{\citenamefont {Chatelin}(2011)}]{chatelin2011spectral}%
  \BibitemOpen
  \bibfield  {author} {\bibinfo {author} {\bibfnamefont {F.}~\bibnamefont
  {Chatelin}},\ }\href {\doibase 10.1137/1.9781611970678} {\emph {\bibinfo
  {title} {Spectral Approximation of Linear Operators}}},\ Classics in Applied
  Mathematics\ (\bibinfo  {publisher} {SIAM},\ \bibinfo {address}
  {Philadelphia},\ \bibinfo {year} {2011})\BibitemShut {NoStop}%
\bibitem [{\citenamefont {Gorissen}\ and\ \citenamefont
  {Vanderzande}(2011)}]{gorissen2011}%
  \BibitemOpen
  \bibfield  {author} {\bibinfo {author} {\bibfnamefont {M.}~\bibnamefont
  {Gorissen}}\ and\ \bibinfo {author} {\bibfnamefont {C.}~\bibnamefont
  {Vanderzande}},\ }\bibfield  {title} {\enquote {\bibinfo {title} {Finite size
  scaling of current fluctuations in the totally asymmetric exclusion
  process},}\ }\href {http://stacks.iop.org/1751-8121/44/i=11/a=115005}
  {\bibfield  {journal} {\bibinfo  {journal} {J. Phys. A: Math. Theor.}\
  }\textbf {\bibinfo {volume} {44}},\ \bibinfo {pages} {115005} (\bibinfo
  {year} {2011})}\BibitemShut {NoStop}%
\bibitem [{\citenamefont {Fleming}\ and\ \citenamefont
  {Soner}(2006)}]{fleming2006}%
  \BibitemOpen
  \bibfield  {author} {\bibinfo {author} {\bibfnamefont {W.~H.}\ \bibnamefont
  {Fleming}}\ and\ \bibinfo {author} {\bibfnamefont {H.~M.}\ \bibnamefont
  {Soner}},\ }\href {http://www.springer.com/gp/book/9780387260457} {\emph
  {\bibinfo {title} {Controlled Markov Processes and Viscosity Solutions}}},\
  \bibinfo {series} {Stochastic Modelling and Applied Probability},
  Vol.~\bibinfo {volume} {25}\ (\bibinfo  {publisher} {Springer},\ \bibinfo
  {address} {New York},\ \bibinfo {year} {2006})\BibitemShut {NoStop}%
\bibitem [{\citenamefont {Rousset}(2006)}]{rousset2006control}%
  \BibitemOpen
  \bibfield  {author} {\bibinfo {author} {\bibfnamefont {M.}~\bibnamefont
  {Rousset}},\ }\bibfield  {title} {\enquote {\bibinfo {title} {On the control
  of an interacting particle estimation of {S}chr{\"o}dinger ground states},}\
  }\href {\doibase 10.1137/050640667} {\bibfield  {journal} {\bibinfo
  {journal} {SIAM J. Math. Anal.}\ }\textbf {\bibinfo {volume} {38}},\ \bibinfo
  {pages} {824--844} (\bibinfo {year} {2006})}\BibitemShut {NoStop}%
\bibitem [{\citenamefont {Nemoto}\ \emph {et~al.}(2016)\citenamefont {Nemoto},
  \citenamefont {Bouchet}, \citenamefont {Jack},\ and\ \citenamefont
  {Lecomte}}]{nemoto2016}%
  \BibitemOpen
  \bibfield  {author} {\bibinfo {author} {\bibfnamefont {T.}~\bibnamefont
  {Nemoto}}, \bibinfo {author} {\bibfnamefont {F.}~\bibnamefont {Bouchet}},
  \bibinfo {author} {\bibfnamefont {R.~L.}\ \bibnamefont {Jack}}, \ and\
  \bibinfo {author} {\bibfnamefont {V.}~\bibnamefont {Lecomte}},\ }\bibfield
  {title} {\enquote {\bibinfo {title} {Population-dynamics method with a
  multicanonical feedback control},}\ }\href {\doibase
  10.1103/PhysRevE.93.062123} {\bibfield  {journal} {\bibinfo  {journal} {Phys.
  Rev. E}\ }\textbf {\bibinfo {volume} {93}},\ \bibinfo {pages} {062123}
  (\bibinfo {year} {2016})}\BibitemShut {NoStop}%
\bibitem [{\citenamefont {Nemoto}\ \emph
  {et~al.}(2017{\natexlab{a}})\citenamefont {Nemoto}, \citenamefont {Hidalgo},\
  and\ \citenamefont {Lecomte}}]{nemoto2017}%
  \BibitemOpen
  \bibfield  {author} {\bibinfo {author} {\bibfnamefont {T.}~\bibnamefont
  {Nemoto}}, \bibinfo {author} {\bibfnamefont {E.~Guevara}\ \bibnamefont
  {Hidalgo}}, \ and\ \bibinfo {author} {\bibfnamefont {V.}~\bibnamefont
  {Lecomte}},\ }\bibfield  {title} {\enquote {\bibinfo {title} {Finite-time and
  finite-size scalings in the evaluation of large-deviation functions:
  {A}nalytical study using a birth-death process},}\ }\href {\doibase
  10.1103/PhysRevE.95.012102} {\bibfield  {journal} {\bibinfo  {journal} {Phys.
  Rev. E}\ }\textbf {\bibinfo {volume} {95}},\ \bibinfo {pages} {012102}
  (\bibinfo {year} {2017}{\natexlab{a}})}\BibitemShut {NoStop}%
\bibitem [{\citenamefont {Kloeden}\ and\ \citenamefont
  {Platen}(1992)}]{kloeden1992}%
  \BibitemOpen
  \bibfield  {author} {\bibinfo {author} {\bibfnamefont {P.~E.}\ \bibnamefont
  {Kloeden}}\ and\ \bibinfo {author} {\bibfnamefont {E.}~\bibnamefont
  {Platen}},\ }\href {http://www.springer.com/gb/book/9783540540625} {\emph
  {\bibinfo {title} {Numerical Solution of Stochastic Differential
  Equations}}}\ (\bibinfo  {publisher} {Springer},\ \bibinfo {address}
  {Berlin},\ \bibinfo {year} {1992})\BibitemShut {NoStop}%
\bibitem [{\citenamefont {Demmel}(1997)}]{demmel1997applied}%
  \BibitemOpen
  \bibfield  {author} {\bibinfo {author} {\bibfnamefont {J.}~\bibnamefont
  {Demmel}},\ }\href {\doibase 10.1137/1.9781611971446} {\emph {\bibinfo
  {title} {Applied Numerical Linear Algebra}}}\ (\bibinfo  {publisher} {SIAM},\
  \bibinfo {address} {Philadelphia},\ \bibinfo {year} {1997})\BibitemShut
  {NoStop}%
\bibitem [{\citenamefont {Polyak}\ and\ \citenamefont
  {Juditsky}(1992)}]{polyak1992acceleration}%
  \BibitemOpen
  \bibfield  {author} {\bibinfo {author} {\bibfnamefont {B.~T.}\ \bibnamefont
  {Polyak}}\ and\ \bibinfo {author} {\bibfnamefont {A.~B.}\ \bibnamefont
  {Juditsky}},\ }\bibfield  {title} {\enquote {\bibinfo {title} {Acceleration
  of stochastic approximation by averaging},}\ }\href {\doibase
  10.1137/0330046} {\bibfield  {journal} {\bibinfo  {journal} {SIAM J. Cont.
  Opt.}\ }\textbf {\bibinfo {volume} {30}},\ \bibinfo {pages} {838--855}
  (\bibinfo {year} {1992})}\BibitemShut {NoStop}%
\bibitem [{\citenamefont {Hartmann}\ and\ \citenamefont
  {Sch\"utte}(2012)}]{hartmann2012}%
  \BibitemOpen
  \bibfield  {author} {\bibinfo {author} {\bibfnamefont {C.}~\bibnamefont
  {Hartmann}}\ and\ \bibinfo {author} {\bibfnamefont {C.}~\bibnamefont
  {Sch\"utte}},\ }\bibfield  {title} {\enquote {\bibinfo {title} {Efficient
  rare event simulation by optimal nonequilibrium forcing},}\ }\href {\doibase
  10.1088/1742-5468/2012/11/P11004} {\bibfield  {journal} {\bibinfo  {journal}
  {J. Stat. Mech.}\ }\textbf {\bibinfo {volume} {2012}},\ \bibinfo {pages}
  {P11004} (\bibinfo {year} {2012})}\BibitemShut {NoStop}%
\bibitem [{\citenamefont {Hartmann}\ \emph {et~al.}(2014)\citenamefont
  {Hartmann}, \citenamefont {Banisch}, \citenamefont {Sarich}, \citenamefont
  {Badowski},\ and\ \citenamefont {Sch\"utte}}]{hartmann2014}%
  \BibitemOpen
  \bibfield  {author} {\bibinfo {author} {\bibfnamefont {C.}~\bibnamefont
  {Hartmann}}, \bibinfo {author} {\bibfnamefont {R.}~\bibnamefont {Banisch}},
  \bibinfo {author} {\bibfnamefont {M.}~\bibnamefont {Sarich}}, \bibinfo
  {author} {\bibfnamefont {T.}~\bibnamefont {Badowski}}, \ and\ \bibinfo
  {author} {\bibfnamefont {C.}~\bibnamefont {Sch\"utte}},\ }\bibfield  {title}
  {\enquote {\bibinfo {title} {Characterization of rare events in molecular
  dynamics},}\ }\href {\doibase 10.3390/e16010350} {\bibfield  {journal}
  {\bibinfo  {journal} {Entropy}\ }\textbf {\bibinfo {volume} {16}},\ \bibinfo
  {pages} {350} (\bibinfo {year} {2014})}\BibitemShut {NoStop}%
\bibitem [{\citenamefont {Zhang}\ \emph {et~al.}(2014)\citenamefont {Zhang},
  \citenamefont {Wang}, \citenamefont {Hartmann}, \citenamefont {Weber},\ and\
  \citenamefont {Sch\"utte}}]{zhang2014}%
  \BibitemOpen
  \bibfield  {author} {\bibinfo {author} {\bibfnamefont {W.}~\bibnamefont
  {Zhang}}, \bibinfo {author} {\bibfnamefont {H.}~\bibnamefont {Wang}},
  \bibinfo {author} {\bibfnamefont {C.}~\bibnamefont {Hartmann}}, \bibinfo
  {author} {\bibfnamefont {M.}~\bibnamefont {Weber}}, \ and\ \bibinfo {author}
  {\bibfnamefont {C.}~\bibnamefont {Sch\"utte}},\ }\bibfield  {title} {\enquote
  {\bibinfo {title} {Applications of the cross-entropy method to importance
  sampling and optimal control of diffusions},}\ }\href {\doibase
  10.1137/14096493X} {\bibfield  {journal} {\bibinfo  {journal} {SIAM J. Sci.
  Comp.}\ }\textbf {\bibinfo {volume} {36}},\ \bibinfo {pages} {A2654--A2672}
  (\bibinfo {year} {2014})}\BibitemShut {NoStop}%
\bibitem [{\citenamefont {Rohwer}\ \emph {et~al.}(2015)\citenamefont {Rohwer},
  \citenamefont {Angeletti},\ and\ \citenamefont {Touchette}}]{rohwer2014}%
  \BibitemOpen
  \bibfield  {author} {\bibinfo {author} {\bibfnamefont {C.~M.}\ \bibnamefont
  {Rohwer}}, \bibinfo {author} {\bibfnamefont {F.}~\bibnamefont {Angeletti}}, \
  and\ \bibinfo {author} {\bibfnamefont {H.}~\bibnamefont {Touchette}},\
  }\bibfield  {title} {\enquote {\bibinfo {title} {Convergence of large
  deviation estimators},}\ }\href {\doibase 10.1103/PhysRevE.92.052104}
  {\bibfield  {journal} {\bibinfo  {journal} {Phys. Rev. E}\ }\textbf {\bibinfo
  {volume} {92}},\ \bibinfo {pages} {052104} (\bibinfo {year}
  {2015})}\BibitemShut {NoStop}%
\bibitem [{\citenamefont {Nemoto}\ and\ \citenamefont
  {Sasa}(2014)}]{nemoto2014}%
  \BibitemOpen
  \bibfield  {author} {\bibinfo {author} {\bibfnamefont {T.}~\bibnamefont
  {Nemoto}}\ and\ \bibinfo {author} {\bibfnamefont {S.-I.}\ \bibnamefont
  {Sasa}},\ }\bibfield  {title} {\enquote {\bibinfo {title} {Computation of
  large deviation statistics via iterative measurement-and-feedback
  procedure},}\ }\href {\doibase 10.1103/PhysRevLett.112.090602} {\bibfield
  {journal} {\bibinfo  {journal} {Phys. Rev. Lett.}\ }\textbf {\bibinfo
  {volume} {112}},\ \bibinfo {pages} {090602} (\bibinfo {year}
  {2014})}\BibitemShut {NoStop}%
\bibitem [{\citenamefont {Risken}(1996)}]{risken1996}%
  \BibitemOpen
  \bibfield  {author} {\bibinfo {author} {\bibfnamefont {H.}~\bibnamefont
  {Risken}},\ }\href
  {http://books.google.co.uk/books?hl=en&lr=&id=MG2V9vTgSgEC&oi=fnd&pg=PA1&dq=risken&ots=dXVCcilAD-&sig=5B2cslOPyHXgNRvD4eTODTm1c-Y#v=onepage&q&f=false}
  {\emph {\bibinfo {title} {The {F}okker-{P}lanck {E}quation: {M}ethods of
  {S}olution and {A}pplications}}},\ \bibinfo {edition} {3rd}\ ed.\ (\bibinfo
  {publisher} {Springer},\ \bibinfo {address} {Berlin},\ \bibinfo {year}
  {1996})\BibitemShut {NoStop}%
\bibitem [{\citenamefont {Reimann}(2002)}]{reimann2002}%
  \BibitemOpen
  \bibfield  {author} {\bibinfo {author} {\bibfnamefont {P.}~\bibnamefont
  {Reimann}},\ }\bibfield  {title} {\enquote {\bibinfo {title} {Brownian
  motors: {N}oisy transport far from equilibrium},}\ }\href {\doibase
  10.1016/S0370-1573(01)00081-3} {\bibfield  {journal} {\bibinfo  {journal}
  {Phys. Rep.}\ }\textbf {\bibinfo {volume} {361}},\ \bibinfo {pages} {57--265}
  (\bibinfo {year} {2002})}\BibitemShut {NoStop}%
\bibitem [{\citenamefont {Ciliberto}\ \emph {et~al.}(2010)\citenamefont
  {Ciliberto}, \citenamefont {Joubaud},\ and\ \citenamefont
  {Petrosyan}}]{ciliberto2010}%
  \BibitemOpen
  \bibfield  {author} {\bibinfo {author} {\bibfnamefont {S.}~\bibnamefont
  {Ciliberto}}, \bibinfo {author} {\bibfnamefont {S.}~\bibnamefont {Joubaud}},
  \ and\ \bibinfo {author} {\bibfnamefont {A.}~\bibnamefont {Petrosyan}},\
  }\bibfield  {title} {\enquote {\bibinfo {title} {Fluctuations in
  out-of-equilibrium systems: {F}rom theory to experiment},}\ }\href {\doibase
  10.1088/1742-5468/2010/12/P12003} {\bibfield  {journal} {\bibinfo  {journal}
  {J. Stat. Mech.}\ }\textbf {\bibinfo {volume} {2010}},\ \bibinfo {pages}
  {P12003} (\bibinfo {year} {2010})}\BibitemShut {NoStop}%
\bibitem [{\citenamefont {{Tsobgni Nyawo}}\ and\ \citenamefont
  {Touchette}(2016)}]{tsobgni2016}%
  \BibitemOpen
  \bibfield  {author} {\bibinfo {author} {\bibfnamefont {P.}~\bibnamefont
  {{Tsobgni Nyawo}}}\ and\ \bibinfo {author} {\bibfnamefont {H.}~\bibnamefont
  {Touchette}},\ }\bibfield  {title} {\enquote {\bibinfo {title} {Large
  deviations of the current for driven periodic diffusions},}\ }\href {\doibase
  10.1103/PhysRevE.94.032101} {\bibfield  {journal} {\bibinfo  {journal} {Phys.
  Rev. E}\ }\textbf {\bibinfo {volume} {94}},\ \bibinfo {pages} {032101}
  (\bibinfo {year} {2016})}\BibitemShut {NoStop}%
\bibitem [{\citenamefont {Mehl}\ \emph {et~al.}(2008)\citenamefont {Mehl},
  \citenamefont {Speck},\ and\ \citenamefont {Seifert}}]{mehl2008}%
  \BibitemOpen
  \bibfield  {author} {\bibinfo {author} {\bibfnamefont {J.}~\bibnamefont
  {Mehl}}, \bibinfo {author} {\bibfnamefont {T.}~\bibnamefont {Speck}}, \ and\
  \bibinfo {author} {\bibfnamefont {U.}~\bibnamefont {Seifert}},\ }\bibfield
  {title} {\enquote {\bibinfo {title} {Large deviation function for entropy
  production in driven one-dimensional systems},}\ }\href {\doibase
  10.1103/PhysRevE.78.011123} {\bibfield  {journal} {\bibinfo  {journal} {Phys.
  Rev. E}\ }\textbf {\bibinfo {volume} {78}},\ \bibinfo {pages} {011123}
  (\bibinfo {year} {2008})}\BibitemShut {NoStop}%
\bibitem [{\citenamefont {Nemoto}\ and\ \citenamefont
  {Sasa}(2011)}]{nemoto2011b}%
  \BibitemOpen
  \bibfield  {author} {\bibinfo {author} {\bibfnamefont {T.}~\bibnamefont
  {Nemoto}}\ and\ \bibinfo {author} {\bibfnamefont {S.-I.}\ \bibnamefont
  {Sasa}},\ }\bibfield  {title} {\enquote {\bibinfo {title} {Variational
  formula for experimental determination of high-order correlations of current
  fluctuations in driven systems},}\ }\href {\doibase
  10.1103/PhysRevE.83.030105} {\bibfield  {journal} {\bibinfo  {journal} {Phys.
  Rev. E}\ }\textbf {\bibinfo {volume} {83}},\ \bibinfo {pages} {030105}
  (\bibinfo {year} {2011})}\BibitemShut {NoStop}%
\bibitem [{\citenamefont {Dupuis}\ and\ \citenamefont
  {Wang}(2005)}]{dupuis2005}%
  \BibitemOpen
  \bibfield  {author} {\bibinfo {author} {\bibfnamefont {P.}~\bibnamefont
  {Dupuis}}\ and\ \bibinfo {author} {\bibfnamefont {H.}~\bibnamefont {Wang}},\
  }\bibfield  {title} {\enquote {\bibinfo {title} {Dynamic importance sampling
  for uniformly recurrent {M}arkov chains},}\ }\href
  {http://www.jstor.org/stable/30038306} {\bibfield  {journal} {\bibinfo
  {journal} {Ann. Appl. Prob.}\ }\textbf {\bibinfo {volume} {15}},\ \bibinfo
  {pages} {1--38} (\bibinfo {year} {2005})}\BibitemShut {NoStop}%
\bibitem [{\citenamefont {Nemoto}\ \emph
  {et~al.}(2017{\natexlab{b}})\citenamefont {Nemoto}, \citenamefont {Jack},\
  and\ \citenamefont {Lecomte}}]{nemoto2017b}%
  \BibitemOpen
  \bibfield  {author} {\bibinfo {author} {\bibfnamefont {T.}~\bibnamefont
  {Nemoto}}, \bibinfo {author} {\bibfnamefont {R.~L.}\ \bibnamefont {Jack}}, \
  and\ \bibinfo {author} {\bibfnamefont {V.}~\bibnamefont {Lecomte}},\
  }\bibfield  {title} {\enquote {\bibinfo {title} {Finite-size scaling of a
  first-order dynamical phase transition: {A}daptive population dynamics and an
  effective model},}\ }\href {\doibase 10.1103/PhysRevLett.118.115702}
  {\bibfield  {journal} {\bibinfo  {journal} {Phys. Rev. Lett.}\ }\textbf
  {\bibinfo {volume} {118}},\ \bibinfo {pages} {115702} (\bibinfo {year}
  {2017}{\natexlab{b}})}\BibitemShut {NoStop}%
\bibitem [{\citenamefont {Blanchet}\ and\ \citenamefont
  {Lam}(2012)}]{blanchet2012}%
  \BibitemOpen
  \bibfield  {author} {\bibinfo {author} {\bibfnamefont {J.}~\bibnamefont
  {Blanchet}}\ and\ \bibinfo {author} {\bibfnamefont {H.}~\bibnamefont {Lam}},\
  }\bibfield  {title} {\enquote {\bibinfo {title} {State-dependent importance
  sampling for rare-event simulation: {A}n overview and recent advances},}\
  }\href {\doibase https://doi.org/10.1016/j.sorms.2011.09.002} {\bibfield
  {journal} {\bibinfo  {journal} {Surv. Oper. Res. Manag. Sci.}\ }\textbf
  {\bibinfo {volume} {17}},\ \bibinfo {pages} {38--59} (\bibinfo {year}
  {2012})}\BibitemShut {NoStop}%
\bibitem [{\citenamefont {Kappen}\ and\ \citenamefont
  {Ruiz}(2016)}]{kappen2016}%
  \BibitemOpen
  \bibfield  {author} {\bibinfo {author} {\bibfnamefont {H.~J.}\ \bibnamefont
  {Kappen}}\ and\ \bibinfo {author} {\bibfnamefont {H.~C.}\ \bibnamefont
  {Ruiz}},\ }\bibfield  {title} {\enquote {\bibinfo {title} {Adaptive
  importance sampling for control and inference},}\ }\href {\doibase
  10.1007/s10955-016-1446-7} {\bibfield  {journal} {\bibinfo  {journal} {J.
  Stat. Phys.}\ }\textbf {\bibinfo {volume} {162}},\ \bibinfo {pages}
  {1244--1266} (\bibinfo {year} {2016})}\BibitemShut {NoStop}%
\bibitem [{\citenamefont {Foulkes}\ \emph {et~al.}(2001)\citenamefont
  {Foulkes}, \citenamefont {Mitas}, \citenamefont {Needs},\ and\ \citenamefont
  {Rajagopal}}]{foulkes2001quantum}%
  \BibitemOpen
  \bibfield  {author} {\bibinfo {author} {\bibfnamefont {W.~M.~C.}\
  \bibnamefont {Foulkes}}, \bibinfo {author} {\bibfnamefont {L.}~\bibnamefont
  {Mitas}}, \bibinfo {author} {\bibfnamefont {R.~J.}\ \bibnamefont {Needs}}, \
  and\ \bibinfo {author} {\bibfnamefont {G.}~\bibnamefont {Rajagopal}},\
  }\bibfield  {title} {\enquote {\bibinfo {title} {Quantum {M}onte {C}arlo
  simulations of solids},}\ }\href {\doibase 10.1103/RevModPhys.73.33}
  {\bibfield  {journal} {\bibinfo  {journal} {Rev. Mod. Phys.}\ }\textbf
  {\bibinfo {volume} {73}},\ \bibinfo {pages} {33} (\bibinfo {year}
  {2001})}\BibitemShut {NoStop}%
\bibitem [{\citenamefont {Lim}\ and\ \citenamefont
  {Weare}(2017)}]{lim2017fast}%
  \BibitemOpen
  \bibfield  {author} {\bibinfo {author} {\bibfnamefont {L.-H.}\ \bibnamefont
  {Lim}}\ and\ \bibinfo {author} {\bibfnamefont {J.}~\bibnamefont {Weare}},\
  }\bibfield  {title} {\enquote {\bibinfo {title} {Fast randomized iteration:
  {D}iffusion {M}onte {C}arlo through the lens of numerical linear algebra},}\
  }\href {\doibase 10.1137/15M1040827} {\bibfield  {journal} {\bibinfo
  {journal} {SIAM Rev.}\ }\textbf {\bibinfo {volume} {59}},\ \bibinfo {pages}
  {547--587} (\bibinfo {year} {2017})}\BibitemShut {NoStop}%
\end{thebibliography}%
\end{document}